\documentclass[10pt, conf, twocolumn]{IEEEtran}

\usepackage{cleveref}
\usepackage{color}
\usepackage{times}
\usepackage{epsfig}
\usepackage{graphicx}
\usepackage{epstopdf}
\usepackage{algorithm}
\usepackage{algorithmic}
\usepackage{amsmath}
\usepackage{amssymb}
\usepackage{bbm}
\usepackage{amsxtra}
\usepackage{multirow}
\usepackage{mathtools}
\usepackage{caption}
\usepackage{cleveref}
\usepackage{url}
\usepackage{float}
\usepackage[font=small, singlelinecheck=false]{caption}
\usepackage{caption}
\usepackage[caption = false]{subfig}
\usepackage{cite}
\usepackage{amsthm}

\newtheorem*{proof*}{Proof}

\begin{document}

\title{A Multi-Layer Feedback System Approach to Resilient Connectivity of Remotely Deployed Mobile Internet of Things}

\author{ \IEEEauthorblockN{\large Muhammad Junaid Farooq, \textit{Student Member, IEEE}\\ and Quanyan Zhu}, \textit{Member, IEEE} \vspace{-0.0in}
%{\thanks {\vspace{-0.2cm}\hrule \vspace{0.2cm} \indent This work was made possible by.}}
\thanks{\vspace{-0.2in}\hrule \vspace{0.2cm}
A part of this work has been presented at the IEEE Global Communications Conference (Globecom 2017), Singapore, Dec. 2017.~\cite{junaid_globecom}.

This research is partially supported by a DHS grant through Critical Infrastructure Resilience Institute (CIRI), grants CNS-1544782 and SES-1541164 from National Science of Foundation (NSF), and grant DE-NE0008571 from the Department of Energy (DOE). The statements made
herein are solely the responsibility of the authors.

%This work was made possible by grant ----- from the -------. The statements made herein are solely the responsibility of the authors.

Muhammad Junaid Farooq and Quanyan Zhu are with the Department of Electrical \& Computer Engineering, Tandon School of Engineering, New York University, Brooklyn, NY, USA, E-mails: \{mjf514, qz494\}@nyu.edu.
}

}

\maketitle

\begin{abstract}
Enabling the Internet of things in remote environments without traditional communication infrastructure requires a multi-layer network architecture. Devices in the overlay network such as unmanned aerial vehicles (UAVs) are required to provide coverage to underlay devices as well as remain connected to other overlay devices to exploit device-to-device (D2D) communication. The coordination, planning, and design of such overlay networks constrained by the underlay devices is a challenging problem. Existing frameworks for placement of UAVs do not consider the lack of backhaul connectivity and the need for D2D communication. Furthermore, they ignore the dynamical aspects of connectivity in such networks which presents additional challenges. For instance, the connectivity of devices can be affected by changes in the network, e.g., the mobility of underlay devices or unavailability of overlay devices due to failure or adversarial attacks. To this end, this work proposes a feedback based adaptive, self-configurable, and resilient framework for the overlay network that cognitively adapts to the changes in the network to provide reliable connectivity between spatially dispersed smart devices. \textcolor{black}{Results show that the proposed framework requires significantly lower number of aerial base stations to provide higher coverage and connectivity to remotely deployed mobile devices as compared to existing approaches.}

\end{abstract}

\IEEEpeerreviewmaketitle

\begin{IEEEkeywords}
Connectivity, feedback, Internet of things, resilience, unmanned aerial vehicles.
\end{IEEEkeywords}

\vspace{-0.0in}
\section{Introduction}
Connectivity between smart devices is vital in enabling the emerging paradigm of the Internet of things (IoT)~\cite{iot_ref}. The fundamental goal of the IoT is to inter-connect smart objects so that they can exchange data and leverage the capabilities of each other to achieve individual and/or network objectives such as high situational awareness, efficiency, accuracy and revenue, etc. This connectivity relies on wireless communication networks which have their limitations based on the communication technologies involved. Existing IoT devices are connected to an access point using wireless personal area network (WPAN) technologies~\cite{iot_tech} such as WiFi, Bluetooth, Zigbee, etc. The access points are in-turn connected to the wired or wireless backhaul networks using wide area network (WAN) technologies~\cite{wan_iot}. The backhaul network enables connectivity and accessibility between things that are geographically separated. However, they may not always be available such as in remote areas~\cite{iort}, disaster struck areas~\cite{disaster}, and battlefields~\cite{iobt_junaid,junaid_iobt_twc}. Unmanned aerial vehicles (UAVs) and mobile ground stations are the most viable candidates for providing connectivity in such situations. \textcolor{black}{For instance, during the hurricane \emph{Harvey}, nearly 95\% of the cellular sites in Rockport, Texas went out of service resulting in nearly a complete communication blackout in the region~\cite{harvey}. In such emergency scenarios, where the traditional communication infrastructure is completely devastated, UAVs can prove to be a promising solution to help create a temporary network and resume connectivity in a short span of time.} Therefore, there is a growing interest towards the use of drones and UAVs as mobile aerial base stations (BSs) to assist existing cellular LTE networks~\cite{drone_placement}, public safety networks~\cite{drone_public_safety}, and intelligent transportation systems~\cite{uav_its}. While this is promising in urban areas where there is high availability of cellular networks which can be used to connect the UAVs to the backhaul, it might not be possible in rural and/or remote regions.

\begin{figure}
  \centering
  \includegraphics[width=3in]{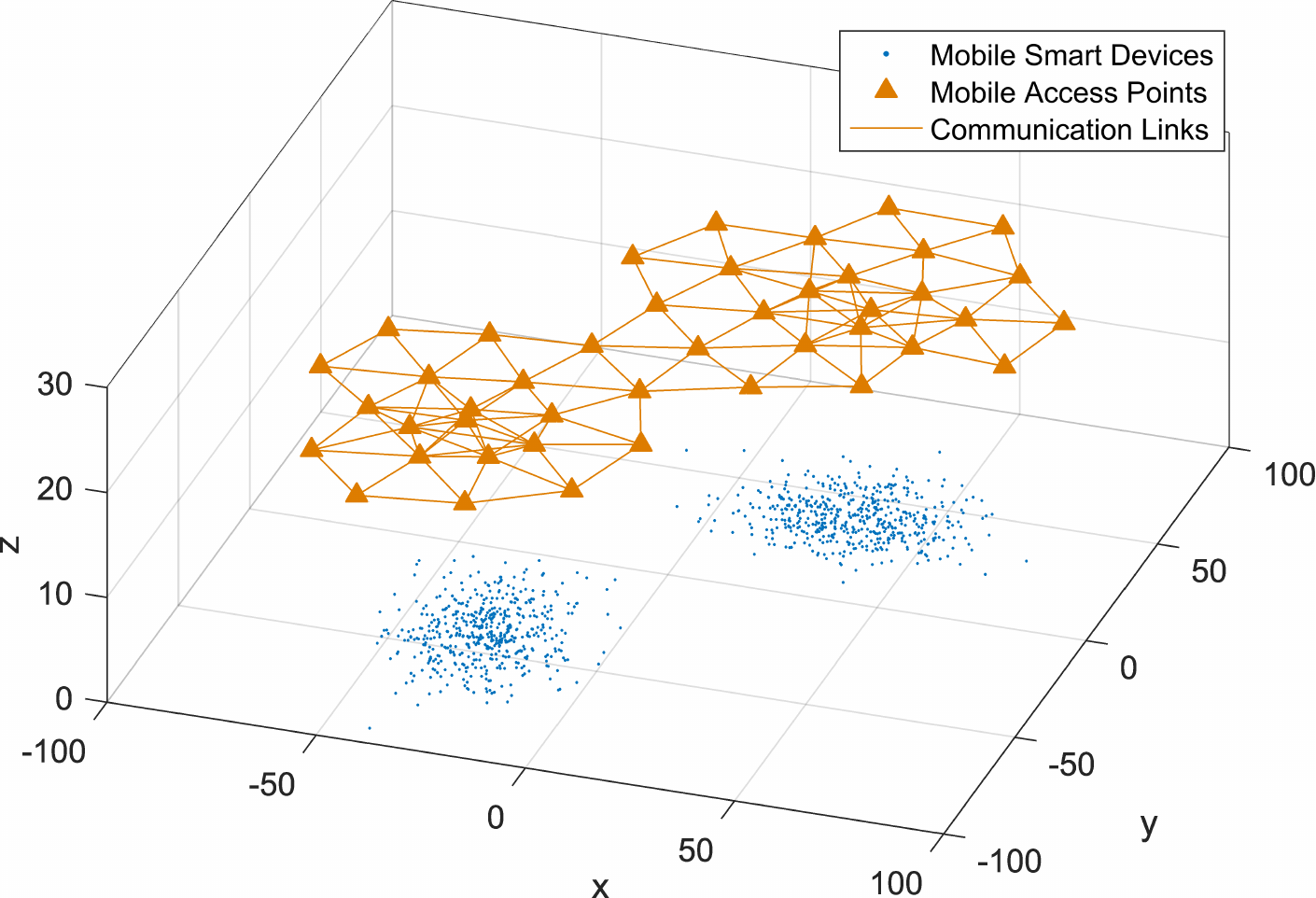}\\
  \caption{Example scenario of spatially clustered mobile smart devices interconnected by an overlay network of mobile access points.\vspace{-0.0in}}\label{remote_iot_fig}
\end{figure}

Due to the absence of traditional communication infrastructure and backhaul networks, the remotely deployed IoT requires a multi-layer architecture comprising of an overlay network of mobile access points (MAPs) to interconnect the spatially dispersed mobile smart devices (MSDs). The MAPs exploit device-to-device (D2D) communications~\cite{d2d_iot,massive_iot} for connecting with other MAPs while the MSDs connect to one of the available MAPs for communication. The problem in such settings is to efficiently deploy the overlay network that provides coverage to all the MSDs as well as maintaining connectivity between the MAPs. Since the MSDs can be located in spatial clusters that are arbitrarily separated, the MAPs should be deployed in a way that they remain connected, i.e., each MAP is reachable from other MAPs using D2D communications. This requirement makes it a challenging network planning and design problem. Fig.~\ref{remote_iot_fig} provides a macroscopic view of one such scenario where the MAPs are appropriately deployed enabling a local inter-network of MSDs without any traditional communication infrastructure. It can be easily connected to the Internet to achieve pervasive connectivity and control over the MSDs. Note that with aerial MAPs, there is an added flexibility to position the BSs arbitrarily in space which might not be possible with other traditional types of access points.

\subsection{Related Work}

There has been an increasing focus recently towards the use of aerial BSs to complement wireless connectivity alongside traditional communication infrastructure~\cite{awais_globecom}. This is due to the significant advancements in drone technology coupled with a massive demand for wireless connectivity with the emergence of the IoT. Several works have considered the use of drone BSs to supplement the coverage of existing cellular networks~\cite{halim2}. In~\cite{UAV_survey}, the authors have provided a review of the opportunities and challenges in using UAVs for wireless communications. Efficient deployment is undoubtedly one of the key challenges in such communication networks.
Therefore, several efforts have been invested in this direction. In~\cite{ot_uav_deployment}, the authors use the classical \emph{optimal transport} framework to obtain a power-efficient deployment of UAVs with the objective to collect data from remotely deployed sensors and not to inter-connect BSs in the air. Similarly, in~\cite{mozaffari_energy_efficient}, an energy efficient UAV placement strategy is developed for the IoT networks. Other examples of works that use a centralized optimization based approach to the UAV placement problem are \cite{shakir_vtc,shakir_pimrc,placement_optimization}. In~\cite{drone_placement}, the authors have proposed a backhaul-aware deployment that is applicable to settings with traditional communication infrastructure. However, in the context of remote IoT, the placement problem is more complex as the BSs have to rely on D2D communication to maintain connectivity. Moreover, in certain applications, it might be inevitable to use such drone-assisted multi-layer architecture.

Most existing works~\cite{ot_uav_deployment,shakir_vtc,shakir_pimrc} dealing with UAV placement formulate the BS deployment problem as a modification of the well-known \emph{facility location} problem, also referred to as the \emph{p-median} problem~\cite{p_median}, from operations research. However, despite being NP-hard, the solution to the facility location problem is not sufficient to ensure the inter-connectivity of the facilities. In our case, the MAPs are wireless devices which have limited communication range and have to be located in sufficient proximity to communicate reliably. Our  goal is to place the MAPs in a connected configuration to enable inter-connectivity between the underlying MSDs using D2D links, which is unique to the wireless network setting. This problem is significantly more complex than the multi-facility location problem. Hence, a globally optimal solution to this problem is not easy to obtain. Moreover, a centralized solution is also less attractive due to the practical limitations in the scenario considered in this paper since the two-layer network cannot be coordinated by a central planner. Another commonly used approach in the literature~\cite{mozaffari_energy_efficient,mozaffari_disk_packing} utilizes the \emph{circle packing}, also referred to as disk packing, solution that aims to cover the area inside a polygon using non-overlapping circles. Again, it is an NP-hard problem and heuristic algorithms exist only for polygon shapes which cannot be extended to arbitrary boundaries making it less attractive for use in the UAV placement frameworks.

%Several efforts have been made in literature towards efficient deployment of aerial BSs to serve ground users such as \cite{drone_placement}.

%In~\cite{circle_packing}, the authors propose a circle packing solution for providing energy efficient coverage in an area. While the circle packing approach does enforce a regular arrangement of the UAVs, the applicability of the solution is limited as it is not scalable and also not well suited for an arbitrary spatial distribution of users.

\subsection{Contributions}
Although the existing approaches provide optimization based approached to the UAV placement problem, we believe that this problem is dynamic in nature and hence a more holistic approach is required to obtain an efficient placement of the MAPs in realtime.
In addition to effective initial deployment of MAPs, there is a need for an autonomic, self-organizing, and self-healing overlay network that can continuously adapt and reconfigure according to the constantly changing network conditions~\cite{chen_paper}. The MSDs can be highly mobile such as smart handheld devices, wireless sensors, and wearable devices whose mobility can be either individual or collective based on the objective such as a rescue operation or a battlefield mission. Furthermore, the network is also vulnerable to failures and cyber-physical adversarial attacks. Therefore, a distributed and dynamic approach to providing resilient connectivity is essential to cope with the growing scale of the networks towards a massive IoT~\cite{massive_iot}. A large body of work is available in the robotics literature dealing with the coverage of a region by autonomous multi-agent systems~\cite{Rus1}. However, they deal primarily with single layer problems and thus the results do not apply directly to a multi-layer network in which one layer aims to provide wireless connectivity to the other. To this end, we propose a feedback based distributed cognitive framework that maintains connectivity of the network and is resilient to the mobility of MSDs and/or failures of the MAPs. The continuous feedback enables the framework to actively react to network changes and appropriately reconfigure the network in response to a failure event that has resulted in loss of connectivity. Simulation results demonstrate that if sufficient MAPs are available, they can be arranged into a desired configuration from arbitrary initial positions and the configuration continuously adapts according to the movement of the MSDs as well as recovers connectivity under varying levels of a random MAP failure event.

The rest of the paper is organized as follows: Section~\ref{Sec:Sys_model} provides the system model describing the connectivity between the overlay and underlay networks. Section~\ref{Sec:Methodology} presents the proposed feedback based cognitive connectivity framework while Section~\ref{Sec:Perf_evaulation} defines the metrics used for performance evaluation. In Section~\ref{Sec:Results}, we provide results on the convergence of the framework and the resilience to mobility and random failures. Finally, Section~\ref{Sec:Conclusion} concludes the paper.

%One related work in literature is the recent development of multi-agent control methods to provide distributed coverage~\cite{coverage_control}. However, the direct application of Voronoi partitioning methods in~\cite{coverage_control} is not viable as the centroids of Voronoi cells may be arbitrarily separated resulting in a disconnected configuration that is not desirable in the D2D network. Another related work is the distributed formation control of autonomous agents that has been developed in the context of robot swarming~\cite{flocking}. One important distinct challenge of our problem is to take into account the spatial distribution of the underlying MSDs as well as the coverage capacity constraints of the overlay MAPs.

%Distributed coverage of underlying users by autonomous agents has been well studied in literature~\cite{coverage_control}. However, it is also based on Voronoi partitioning of the space according to the user density and does not take into account the connectivity of the Voronoi centers. The distributed formation control of autonomous agents in a lattice configuration has been developed in the context of \emph{flocking} of robots~\cite{flocking}. However, it does not take into account the spatial distribution of the underlying network as well as the coverage capacity constraints of the robots.

%D2D backhaul on the fly
\vspace{-0.0in}
\section{System Model}\label{Sec:Sys_model}
\begin{figure}[t]
  \centering
  % Requires \usepackage{graphicx}
  \includegraphics[width=3in]{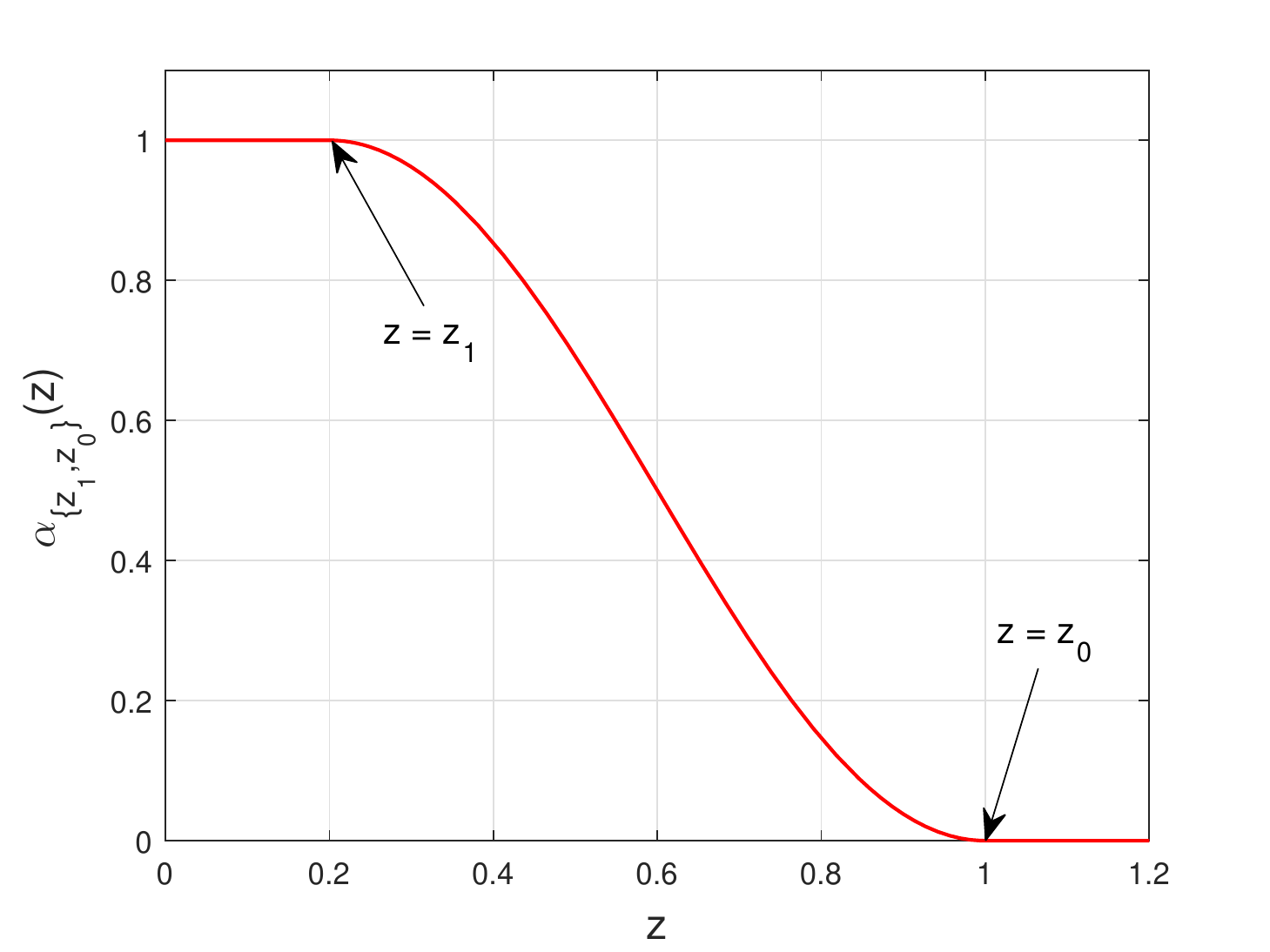}\\
  \caption{Illustration of the strength of communication link function in (1).}\label{bumpfig}
\end{figure}

We consider a finite set of MSDs arbitrarily placed in $\mathbb{R}^2$ denoted by $\mathcal{M} = \{1, \ldots, M\}$ and a finite set of MAPs denoted by $\mathcal{L} = \{1, \ldots, L\}$, placed in $\mathbb{R}^2$ at an elevation of $h\in \mathbb{R}$, for providing connectivity to the MSDs\footnote{1We assume MAPs have a constant elevation from the ground for simplicity,
however, the methodology and results can be readily generalized to varying
elevations.}. The Cartesian coordinates of the MSDs at time $t$ are denoted by $\mathbf{y}(t) = [y_1(t), y_2(t), \ldots, y_M(t)]^T$, where $y_i(t) \in \mathbb{R}^2, \forall i \in \mathcal{M}, t \geq 0$. Similarly, the Cartesian coordinates of the MAPs at time $t$ are denoted by $\mathbf{q}(t) = [q_1(t), q_2(t), \ldots, q_{L}(t)]^T$, where $q_i \in \mathbb{R}^2$, $\forall i \in \mathcal{L}, t \geq 0$. For brevity of notation, we drop the time index henceforth and assume that the time dependence is implicitly implied. \textcolor{black}{Initially, the MSDs are partitioned into $K\in\mathbb{Z}^+$ sets denoted by $\mathcal{S} = \{S_1, S_2, \ldots, S_K\}$.  The centroid of each set or cluster is denoted by $\bar{C}_i$.} The MAPs have a maximum communication range of $r \in \mathbb{R}^+$, i.e., any two MAPs can communicate only if the Euclidean distance between them is less than $r$\footnote{The average communication range can be determined using metrics such as the probability of transmission success together with path loss models developed for air to ground communication channels such as in \cite{PLOS_model,air_to_ground_channel}}. The communication neighbours of each MAP is represented by the set $\mathcal{N}_i = \{ j \in \mathcal{L}, j \neq i : \| q_i - q_j \| \leq r \}, \forall i \in \mathcal{L}$. The quality or strength of the communication links between the MAPs is modeled using a distance dependent decaying function
$\alpha_{\{z_1,z_0\}}(z) \in [0,1]$ with finite cut-offs\footnote{This type of function is referred to as a bump function in Robotics literature. It is pertinent to mention that the cutoff values are crucial while the actual form of the bump function is not important in our application.}, expressed as follows~\cite{bump_ref}:
\begin{align} \label{beta_func}
\alpha_{\{z_1, z_0\}}(z) =
\begin{cases}
1, & \text{if } 0 \leq z < z_1, \\
0.5 \left( 1 + \cos(\pi \frac{z - z_1}{z_0 - z_1}) \right), & \text{if } z_1 \leq z < z_0, \\
0, & \text{if } z \geq z_0,
\end{cases}
\end{align}
where $z_0$ and $z_1$ are constants that define the cut-off values corresponding to $0$ and $1$ respectively. An illustration of the function with $z_1 = 0.2$ and $z_0 = 1$ is shown in Fig.~\ref{bumpfig}.

\begin{figure}[t]
  \centering
  \includegraphics[width=3in]{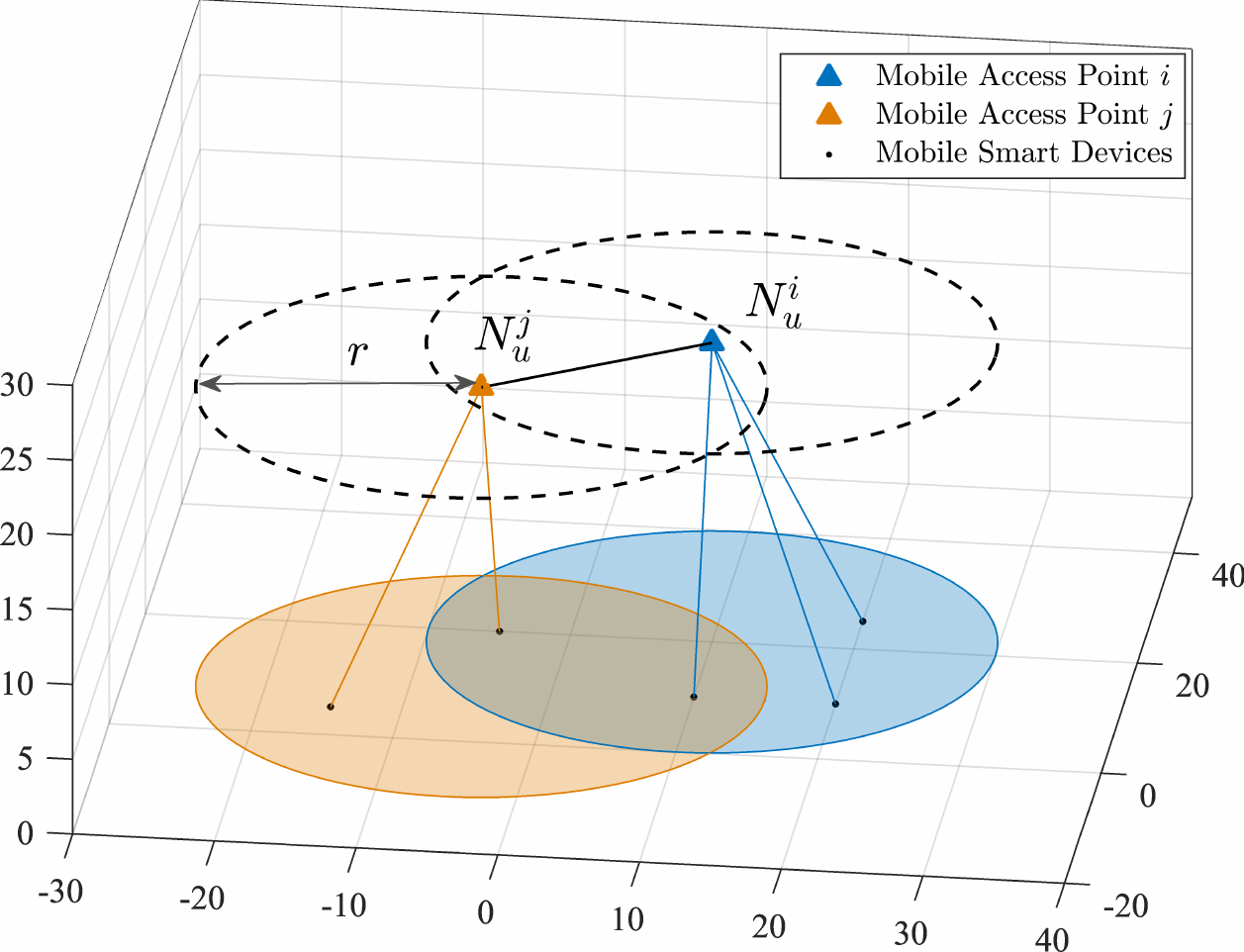}\\
  \caption{An example of two connected MAPs serving the underlying MSDs. nThe communication range of each MAP is depicted by the dotted lines while the area of influence is represented by the shaded circles.}\label{sys_model}
\end{figure}
\textcolor{black}{Note that the function $\alpha_{\{z_1,z_0\}}(z) \in [0,1]$, parameterized by $z_0$ and $z_1$, represents a generic measure of the communication quality between the MAPs with 1 referring to perfect communication and 0 referring to completely absent communication. However, in practice, the constants $z_0$ and $z_1$ need to be carefully selected according to the communication quality requirements based on the physical signal propagation model. One of the possibilities is to use the signal-to-noise-plus-interference-ratio (SINR) to determine the probability of successful transmission between the devices. Thresholds can then be imposed on the coverage probability to obtain the reference distances, which can in turn be normalized to obtain $z_0$ and $z_1$. It is pertinent to mention that obtaining the parameters via an accurate characterization of the success probability based on the SINR results also takes into account the effect of interference in the network. Several analytical approaches such as stochastic geometry can be used to obtain such a characterization~\cite{junaid_vtc, junaid_stochastic}}.
Furthermore, in order to make the norm measure of a vector differentiable at the origin, a new mapping of the \textcolor{black}{$\mathcal{L}_2$} norm is defined following~\cite{flocking}, referred to as the $\sigma-$norm\footnote{Note that this is not a norm but a mapping from a vector space to a scalar.}:
\begin{align}
\|\text{x}\|_{\sigma} = \frac{1}{\epsilon}\left( \sqrt{1 + \epsilon \| \text{x}\|^2} - 1\right),
\end{align}
where $\epsilon > 0$ is a constant. The smooth adjacency matrix containing the strength of linkages between the MAPs, denoted by $\mathbf{A}~=~[a_{ij}] \in \mathbb{R}^{L \times L}$ can then be obtained as follows:
\begin{align}\label{alpha_func}
a_{ij} =
\begin{cases}
\alpha_{\{\gamma,1\}} \left(\frac{ \| q_i - q_j\|_{\sigma}}{\|r\|_{\sigma}} \right), & \text{if } i \neq j,\\
0, & \text{if } i = j.
\end{cases}
\end{align}
%\textcolor{red}{Moreover, we assume that adjacent MAPs use different frequency channels to communicate and hence, do not interfere with each other. Therefore, the reliability of the communication links between the MAPs is directly reflected by the adjacency matrix $\mathbf{A}$.}
Note that the function in \eqref{alpha_func}, uses $z_0 = 1$ and $z_1$ is replaced by an arbitrary parameter $\gamma \in [0,1]$ referring to the upper cutoff value. The degree matrix of the MAPs is defined by $\mathbf{D} = [d_{ij}] \in \mathbb{R}^{L \times L}$, where $d_{ij} = \mathbbm{1}_{i = j} \sum_{j = 1}^{L} \mathbbm{1}_{ \{a_{ij} > 0\}},\\ \forall i,j \in \mathcal{L}$, where $\mathbbm{1}_{\{.\}}$ denotes the indicator function.
The connectivity of MSDs is determined by their coverage by one of the available MAPs.
%A MAP has a certain area of influence under which it can reliably provide coverage to the MSDs. For simplicity, we assume the range of influence of the MAPs on the ground is the same as their communication range.
Fig.~\ref{sys_model} illustrates a typical scenario of two adjacent MAPs providing connectivity to the MSDs inside their influence region enabling network-wide connectivity. An MSD $i$ is assumed to be connected to MAP $j$ if it is closer to it than any other MAP, i.e., $\|y_i - q_j\| < \|y_i - q_k\|, k \in \mathcal{L}\backslash j$, and the MAP has sufficient capacity to serve the MSD. \textcolor{black}{The quality of service or equivalently the utility of the communication link between an MSD $i$ located at $y_i$ and an MAP $j$ located at $q_j$ is represented by:
\begin{align}\label{matching1}
\Phi(i,j) = \phi(\|y_i - q_j\|),
\end{align}
where $\phi: \mathbb{R}^+ \rightarrow \mathbb{R}^+$ is continuously differentiable and decreasing function}.
The total number of MSDs connected to the MAPs are denoted by the vector $\mathbf{N}_u = [N_u^1, N_u^2, \ldots, N_u^L]^T$, while the maximum serving capacity of each MAP is denoted by $N^{\max}$. The distance based user association is motivated by the distance dependent signal decay. Each MSD aspires to connect to its nearest in-rage MAP unless constrained by the capacity of the host.

\vspace{-0.0in}
\section{Methodology} \label{Sec:Methodology}

\begin{figure}
  \centering
  \includegraphics[width=3.4in]{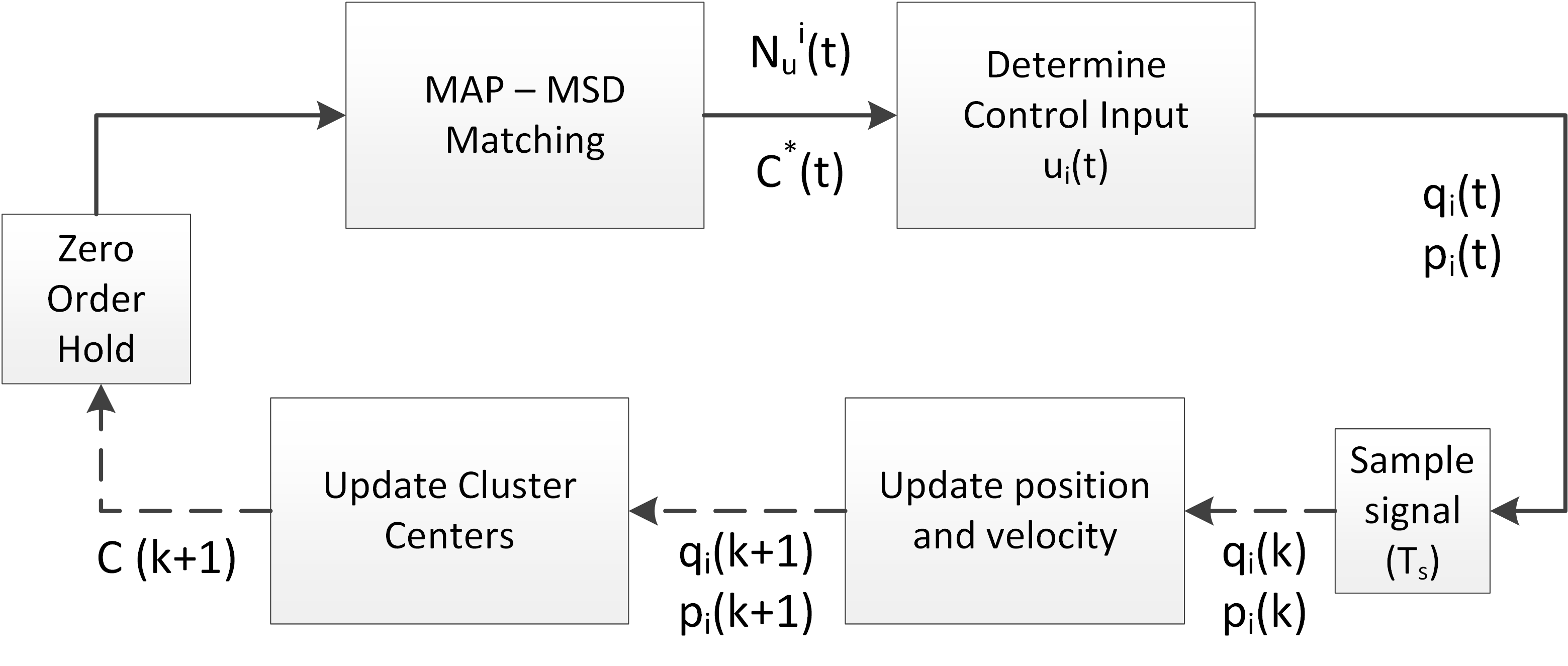}\\
  \caption{Feedback cognitive loop for the proposed resilient connectivity framework. Each MAP computes its control input based on the information about the location of MSDs and the number of connected users. The states of the MAPs are then updated in discrete time before repeating the observation cycle.}\label{cognition}
\end{figure}

In this section, we describe the methodology used to develop the cognitive and resilient connectivity framework for remotely deployed IoT devices. We assume that the locations of the MSDs are constantly changing and is beyond the control of the MAPs. Our objective is to autonomously configure the MAPs in a distributed manner to provide coverage to the MSDs as well as keeping them connected to other MAPs. The cognitive connectivity resilience framework can be summarized by the cognition loop illustrated in Fig.~\ref{cognition}. The individual blocks of the cognitive framework are elaborated in the subsequent subsections.

\subsection{MAP-MSD Matching}
At each iteration of the cognitive connectivity framework, there need to be an association between the MAPs and the MSDs. Since the wireless channel experiences distance dependent path loss, it is reasonable to assume a utility based on the Euclidean distance between the MSD and the MAP as given by~\eqref{matching1}. The preference of MSD $i$ to connect to a MAP can be described as follows:
\begin{align}\label{matching2}
v(i) = \underset{j \in \mathcal{L} : \|y_i - q_j\| < r}{\max} \{\Phi(i,j)\}.
\end{align}
Notice, that only the MSDs that are under the influence of the MAPs are matched and the uncovered MSDs remain un-matched. The optimal assignment results in the matrix $\boldsymbol{\varepsilon}$, where $\varepsilon_{ij} = \mathbbm{1}_{\{j = v(i)\}}, \forall i \in \mathcal{M}, j \in \mathcal{L}$. As a result, the number of MSDs matched to each MAP can be evaluated as $N_u^i = \sum_{j = 1}^{M} \varepsilon_{ij}, \forall i \in \mathcal{L}$.

\subsection{MAP Dynamics and Objective}
We employ the kinematic model for the MAPs, whose dynamics can be written as follows:
\begin{align} \label{dynamics}
\dot{q}_i = p_i,\\ \notag
\dot{p}_i = u_i,
\end{align}
where $q_i, p_i, u_i \in \mathbb{R}^2, i \in \mathcal{L}$, represent the displacement, velocity, and acceleration of the devices respectively. It can be represented using the state space representation $\dot{\mathbf{x}}_i = \tilde{A}_c \mathbf{x}_i + \tilde{B}_c \mathbf{u}_i$, where the vectors and matrices are defined as follows:
\begin{align*}
\mathbf{x}_i = \Bigg[ \ \begin{matrix}q_i \\ p_i\end{matrix} \ \Bigg], \mathbf{u}_i = \Bigg[ \ \begin{matrix}0 \\ u_i\end{matrix} \ \Bigg],  \tilde{A}_c = \Bigg[ \ \begin{matrix}0 & 1\\ 0 & 0\end{matrix} \ \Bigg], \tilde{B}_c = \Bigg[ \ \begin{matrix}0 \\ 1 \end{matrix} \ \Bigg]
\end{align*}
Note that the \textcolor{black}{continuous time dynamical system} is completely controllable. For practical implementation, the displacement and velocity vectors are discretized with a sampling interval of $T_s$. The equivalent discrete time system can be written as follows:
\begin{align}\label{discrete_dynamics}
\mathbf{x}(k+1)= \tilde{A}_d \mathbf{x}(k) + \tilde{B}_d \mathbf{u}(k),
\end{align}
where the matrices $\tilde{A}_d$ and $\tilde{B}_d$ govern the discrete time dynamics of the system.

The goal is to design a control input $u_i$ for each MAP which eventually leads to a desired configuration. To achieve this, we build upon the framework developed in~\cite{flocking} for distributed multi-agent systems and provide modifications which leads to the desired behaviour of MAPs in the context of D2D wireless networks constrained by the underlying MSDs. To enhance spatial coverage, the MAPs should have less coverage overlap and should be spread out while remaining connected to other MAPs. Therefore, we define a minimum distance $0 \leq d < r$ such that two MAPs should not be closer to each other than $d$ unless they are forced to be closely located to serve a higher density of underlying MSDs.

%\subsection{Cognitive Connectivity Algorithm}
\subsection{Controller Design}

We propose a control input $\mathbf{u} = [u_1, u_2, \ldots, u_{L}]^T$ for each of the MAPs in the following form:
\begin{align}
u_i = f_i(\mathbf{q},\mathbf{A},\mathbf{N}_u) + g_i(\mathbf{p},\mathbf{A}) + h_i(\mathbf{q}, \mathbf{p}), \label{control_input}
\end{align}
where $f_i(\mathbf{q},\mathbf{A},\mathbf{N}_u)$ defines the gradient based term based on the attractive and repulsive forces between the MAPs, $g_i(\mathbf{p},\mathbf{A})$ is the velocity matching term that forces neighbouring MAPs to move with the same speed, and $h_i(\mathbf{q}, \mathbf{p})$ is the term defining the individual goals of each of the MAPs. Note that $\mathbf{q}$, $\mathbf{p}$, and $\mathbf{A}$ are functions of time. Such controllers holistically model the control of autonomous agents and are hence widely used in literature for a variety of robot swarming applications.

Each of the terms in~\eqref{control_input} are elaborated as follows:

\subsubsection{Attractive and Repulsive Function}
As highlighted earlier, the MAPs tend to maintain a minimum distance $d$ with other MAPs unless the serving capacity is exceeded. Therefore, a repulsive force is required from MAP $j$ to MAP $i$ if the distance between them is less than $d$ and an attractive force is needed from MAP $i$ to MAP $j$ if MAP $j$ exceeds the capacity of serving MSDs. In effect, MAP $i$ tends to share the load of MAP $j$ if it exceeds capacity in order to provide coverage to all the MSDs. Therefore, the $i^{th}$ element of $f_i(\mathbf{q},\mathbf{A},\mathbf{N}_u)$ can be defined as follows:
\begin{align}
&f_i(\mathbf{q},\mathbf{A},\mathbf{N}_u) = \sum_{j \in \mathcal{N}_i} \Bigg[ \Psi( \| q_j  - q_i \|_{\sigma} )  \ +   \notag \\& \quad  a \left( 1 - \alpha_{\{0,1\}} \left( \frac{\| (N_u^j - N^{\max})^+\|_{\sigma}}{\| N^{\max}\|_{\sigma}} \right) \right)  \Bigg] \mathbf{v}_{ij},i \in \mathcal{L},\label{grad_term}
\end{align}
where $\mathbf{v}_{ij} = \nabla \| q_j - q_i \|_{\sigma}$ represents the vector in the direction going from the MAP at location $q_i$ to the MAP at location $q_j$. The function $\Psi(z)$ is provided as follows:
\begin{align}
\Psi(z) = \alpha_{\{\gamma,1\}} \left( \frac{z}{\|r\|_{\sigma}} \right) \phi(z - \|d\|_{\sigma}),
\end{align}
where $\phi(z) = \frac{1}{2}[(a+b) \frac{(z+c)}{\sqrt{1 + (z+c)^2}} + (a - b)]$ is an un-even sigmoid function with $c = |a-b|/\sqrt{4ab}$ such that $\phi(z) \in (-a,a)$. Notice that the function $\Psi(z)$ is a product of two functions that results in the property that $\Psi(z) \leq 0$ if $z < \|d\|_{\sigma}$ and $\Psi(z) = 0$ otherwise, i.e., it provides a repelling force if MAP $i$ and MAP $j$ are closer than $d$ and is neutral if they are farther than $d$. Therefore, the first term in the multiplier of the gradient in \eqref{grad_term} ensures that the distance between neighbouring MAPs is at least $d$. The second term is related to the attraction between MAPs if the MSDs aspiring to connect to them are beyond their capacity, i.e., $N^{\max}$. The force depends on the number of unserved users $(N_u^j - N^{\max})^+$ normalized by the maximum capacity and is accomplished using the function $a(1 - \alpha_{\{0,1\}}(.))$ that maps from $0$ to $a$, which is nonzero for strictly positive arguments.

\subsubsection{Velocity Consensus Function}
The velocity consensus function enables a matching between the velocities of neighbouring MAPs and is expressed as follows:
\begin{align}
g_i(\mathbf{p}, \mathbf{A}) = \sum_{j \in \mathcal{N}_i} a_{ij}(p_j - p_i), i \in \mathcal{L}.
\end{align}
The function implies that MAP $i$ tends to align its velocity to its neighbours weighted by the strength of the links.
It generates a damping force in the movement of each of the MAP preventing any erratic behaviour of any of them resulting in an unwanted collision. Furthermore,
in the case where MSDs are moving in a coordinated fashion, the velocity consensus function prevents potential disconnections among the MAPs due to the otherwise relative velocity.
\subsubsection{Individual Goal Function}
The individual goal function $h(\mathbf{q}, \mathbf{p})$ is defined as follows:
\begin{align}
h_i(\mathbf{q}, \mathbf{p}) = c_1 (q_i^r - q_i ) +  c_2 (p_i^r - p_i), i \in \mathcal{L},
\end{align}
where $q_i^r$ and $p_i^r$ are the reference position and velocity of MAP $i$, and $c_1$ and $c_2$ are positive constants denoting the relative aggressiveness to achieve the goal. It serves as a navigational feedback term which determines the eventual state that the MAPs aspire to achieve. If the goal of each MAP is precisely determined, the network can be made to reach the desired configuration. Assuming that each MAP is greedy to serve MSDs, a natural goal is to reach the centroid of the MSDs to allow a maximum number of MSDs to connect to it. Since the MSDs may be arbitrarily clustered, it is more efficient for the MAPs to move toward the centroid that is nearest to them. Hence the individual reference signals are selected as follows:
\begin{align}
q_i^r  =  C_i^*, \ \forall i \in \mathcal{L}, \notag\\
p_i^r  =  0, \ \forall i \in \mathcal{L},
\end{align}
where $C_i^*$ denotes the coordinates of the cluster center nearest to MSD $i$ and is further elaborated in the sequel. A reference velocity of $0$ implies that each MAP wants to eventually come to rest. The proposed cognitive loop propagates as follows: Given the spatial locations of the MSDs, a matching is made between the MAPs and the MSDs based on the distances and the maximum capacity of the MAPs. Based on the MAP-MSD association and the cluster centers (determined by the spatial location of MSDs), a control input is computed by each MAP independently and the system states are updated according to the dynamics provided by \eqref{dynamics}. After the locations and velocities are updated, new cluster centers of the MSDs are computed as their spatial locations might have changed due to mobility. Upon convergence, the control input becomes nearly zero and there is no further change in the configuration provided the MSDs do not change their positions.

\subsection{Cluster Centers}
In order to determine the destination of each MAP, we need information about the locations of the MSDs. Since the MSDs can move arbitrarily, they may not have a definitive spatial distribution. Therefore, it is reasonable to cluster the MSDs and use their centers as a potential destination for nearby MAPs. At each step, the objective is to find the following:
\begin{align}
\underset{\mathcal{S}}{\arg \min} \sum_{i=1}^{K} \sum_{y \in S_i} \| y - \bar{C}_i\|^2,
\end{align}
The resulting optimal cluster centers of $\mathcal{S}$ are denoted by $\mathcal{C} = [C_1, C_2, \ldots, C_K]^T, C_i \in \mathbb{R}^2, \forall i = 1, \ldots, K$.
The solution to this problem can be obtained efficiently using Lloyd's algorithm~\cite{lloyds}. In our proposed model, the only centralized information needed is the coordinates of the cluster centers of the MSDs. It can either be obtained using a centralized entity such as the satellite or it can be locally estimated based on individual observations. However, if local observations are used, then a distributed consensus needs to be made regarding the final goal state of each MAP. Due to the additional complexity in distributed consensus development, we have postponed it for future work. However, interested readers are directed to~\cite{consensus_multi_agent} and~\cite{Fierro1} for more details.
%It remains to be investigated if locally estimated cluster centers by each MAP results in a desirable network formation and is intended as a future work.

The flow of the cognitive connectivity algorithm is formally summarized in Algorithm~\ref{Algorithm1}. At epoch, each MAP obtains information about the centroid of the users known as the cluster centers and determines the coordinates of its nearest centroid denoted by $C_i^*$. Each of the MAPs interact with the users in their area of influence and agree on serving the selected MSDs. The number of users supported by each MAP at time $t$ is denoted by $N_{\text{u}}^i(t)$. Once the MAPs know their individual information, they exchange the information about the position, velocity, and number of connected users with their immediate neighbours. Using the aggregated information, each MAP computes its control input $\mathbf{u}_i$. The position and velocity are then updated using the discrete time state space model in~\eqref{discrete_dynamics} with a time step of $\Delta$. This process is repeatedly executed until the operation of MAPs is terminated. In the subsequent section, we describe the key metrics that are used to evaluate the performance of the proposed cognitive algorithm.
\begin{algorithm}
\caption{Resilient Connectivity Algorithm}
\label{Algorithm1}
\begin{algorithmic}[1]
\REQUIRE Initial position and velocity of each MAP $q(0)$ and $p(0)$; Initial $K$ centroids of the MSDs $\mathcal{C}(0)$; Time step $\Delta$.
\REPEAT \label{loop_start}
\STATE{Determine the coordinates of the nearest cluster center (goal position) for each MAP.}
\STATE{Obtain a matching between each MAP and the MSDs in its area of influence using~\eqref{matching1} and~\eqref{matching2}.}
\STATE{Determine the number of connected users of each MAP, $N_{\text{u}}^{i}(t)$.}
\STATE{Each MAP shares the position, velocity, and number of connected MSDs information with its immediate communication neighbours.}
\STATE{Using $\mathcal{C}(t)$ and $N_{\text{u}}^{i}(t)$, determine control input for each MAP $\mathbf{u}_i(t), i \in \mathcal{L}$ from~\eqref{control_input}.}
\STATE{Update the position and velocity of each MAP using the discretized state space model $\mathbf{x}(k+1)= \tilde{A} \mathbf{x}(k) + \tilde{B} \mathbf{u}(k)$ with a time step of $\Delta$.}
\UNTIL{End of operation.} \label{loop_end}
\end{algorithmic}
\end{algorithm}

\vspace{-0.0in}
\section{Performance Evaluation}\label{Sec:Perf_evaulation}
In order to evaluate the performance of the proposed cognitive connectivity framework, we use the following metrics to measure the connectivity of the network:
\begin{enumerate}
\item \textbf{Proportion of MSDs covered:} It measures the percentage of total MSDs that are associated and served by one of the MAPs. This metric helps in determining the general accessibility of the MSDs to the MAPs. However, it does not reflect the true connectivity of the MSDs as the MAPs might not be well connected.
\item \textbf{Probability of information penetration in MAPs:} Assuming that the connected MSDs can communicate perfectly with the MAPs, the overall performance of the system depends on the effectiveness of communication between the MAPs using D2D links. One way to study the dynamic information propagation and penetration in D2D networks is based on mathematical epidemiology (See~\cite{iobt_junaid}). Since the network in this paper is finite, we make use of the $N$-intertwined mean field epidemic model~\cite{N_intertwined} to characterize the spread of information. The steady state probability of MAP $i$ being informed by a message, denoted by $\nu_{i \infty}$, propagated in the D2D network at an effective spreading rate of $\tau$ is bounded as follows (See Theorem 1 in~\cite{N_intertwined}):
    \begin{align}
    0 \leq \nu_{i \infty} \leq  1 - \frac{1}{1 + \tau d_{ii}}.
    \end{align}
    It measures the probability of a device being informed about a piece of information that initiates from any device in the network at random. In other words, a 70\% probability of information dissemination reflects that if a piece of information is generated at random by any of the nodes, there is a 70\% probability on average that any other node in the network will receive it assuming perfect success in the transmissions.
\item \textbf{Reachability of MAPs:} While it is important to determine the spreading of information over the D2D network, it is also crucial to know whether the D2D network is connected or not. It can be effectively determined using the algebraic connectivity measure from graph theory, also referred to as the \emph{Fiedler value}, i.e., $\lambda_2(\mathbf{L})$, where $\lambda_2(.)$ denotes the second-smallest eigenvalue and $\mathbf{L}$ is the Laplacian matrix of the graph defined by the adjacency matrix $\mathbf{A}$. A nonzero Fiedler value indicates that each MAP in the network is reachable from any of the other MAPs. The Laplacian matrix is defined as follows:
\begin{align}
\mathbf{L} = \mathbf{D} - \mathbf{A}.
\end{align}
\end{enumerate}
These metrics complement each other in understanding the connectivity of the network. The resilience, on the other hand, is measured in terms of the percentage of performance recovery after an event of failure has occurred. In the following subsection, we provide the simulations results and describe the main observations in comparison with existing approaches.
%Information spread in contact based networks can be modeled by the epidemic models used for studying the spread of viruses.
%How effectively can information be propagated into the network

\begin{table}[]
\centering
\caption{Simulation Parameters.}
\label{parameters}
\begin{tabular}{|l|l|l|l|}
\hline
Parameter & Value & Parameter & Value \\ \hline
$M$         & 2000  & $a$         & 5     \\ \hline
$L$         & 80    & $b$         & 5     \\ \hline
$r$         & 24    & $c_1$        & 0.2   \\ \hline
$d$         & 20    & $c_2$        & 0.1   \\ \hline
$\epsilon$         & 0.1   & $s$         & 0.2   \\ \hline
$N^{\max}$      & 80    &    $\tau$       &  1     \\ \hline
$h$      & 20    &       $T_s$    &   0.01   \\ \hline
$k$      & 3    &    $\gamma$      &    0.2  \\ \hline
\end{tabular}
\end{table}

\begin{figure*}[t!]
\centering
\subfloat[]{\includegraphics[width = 2.2in]{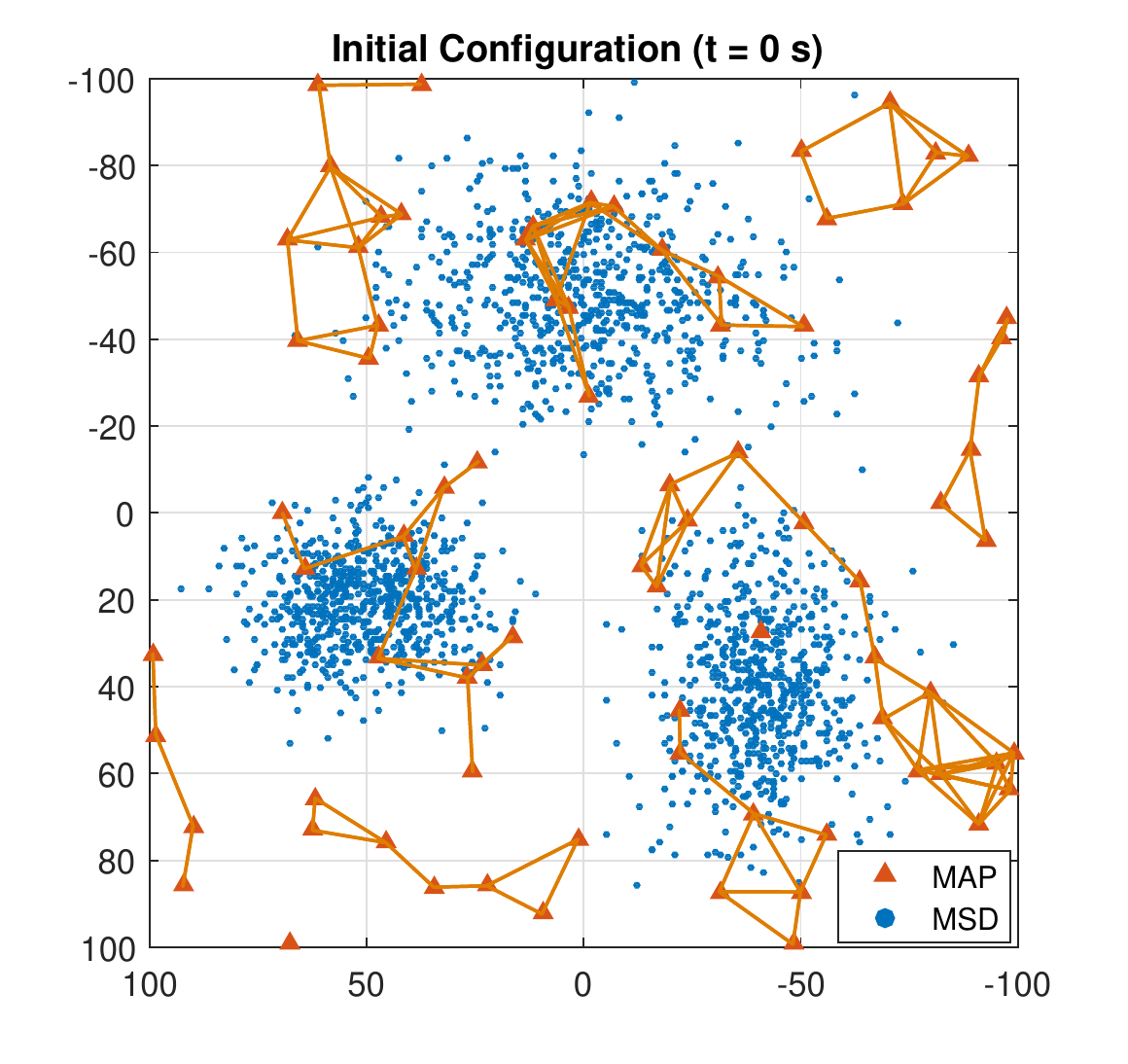} \label{config_t0}}
\subfloat[]{\includegraphics[width = 2.2in]{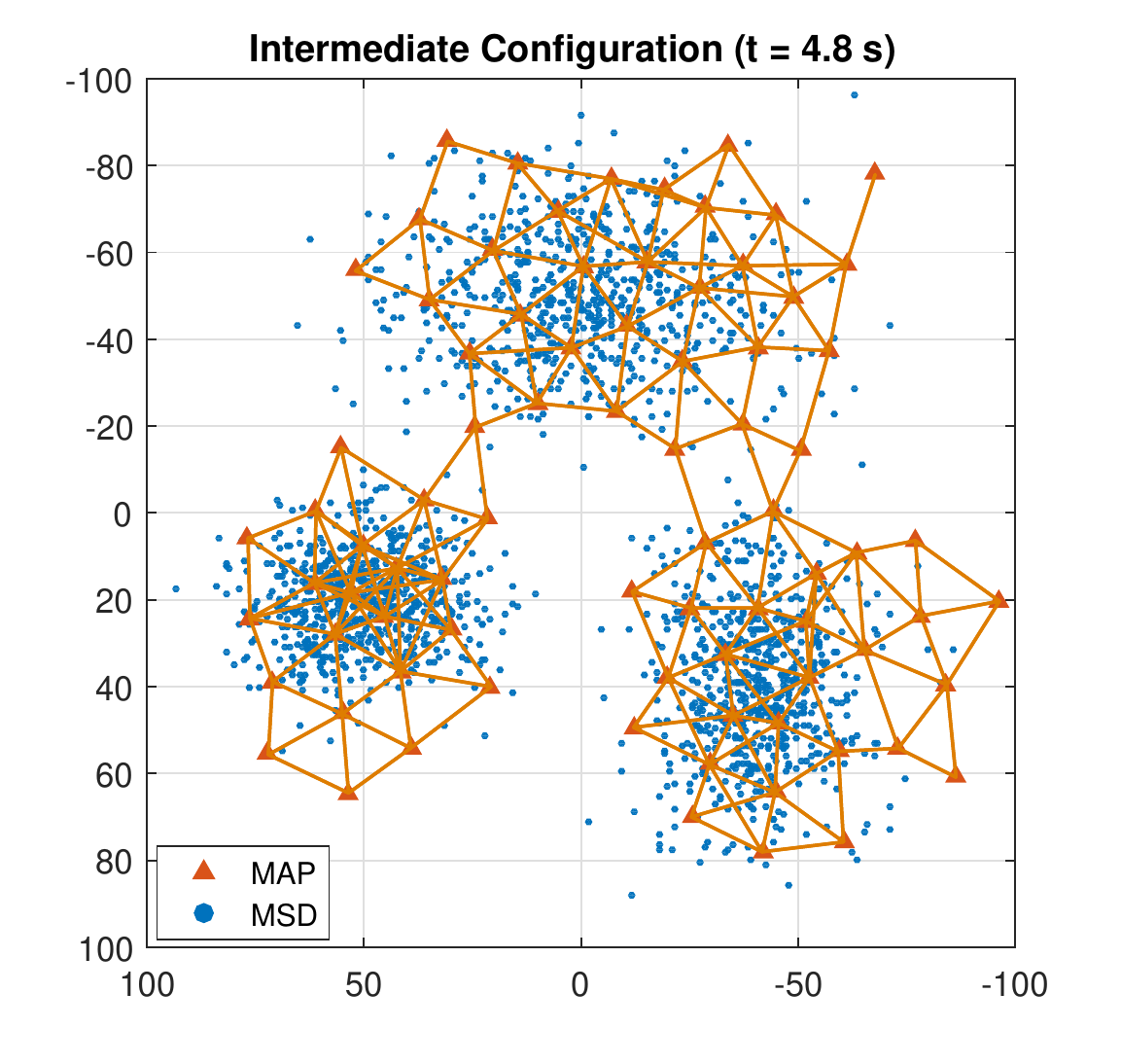} \label{config_t4_8}}
\subfloat[]{\includegraphics[width = 2.2in]{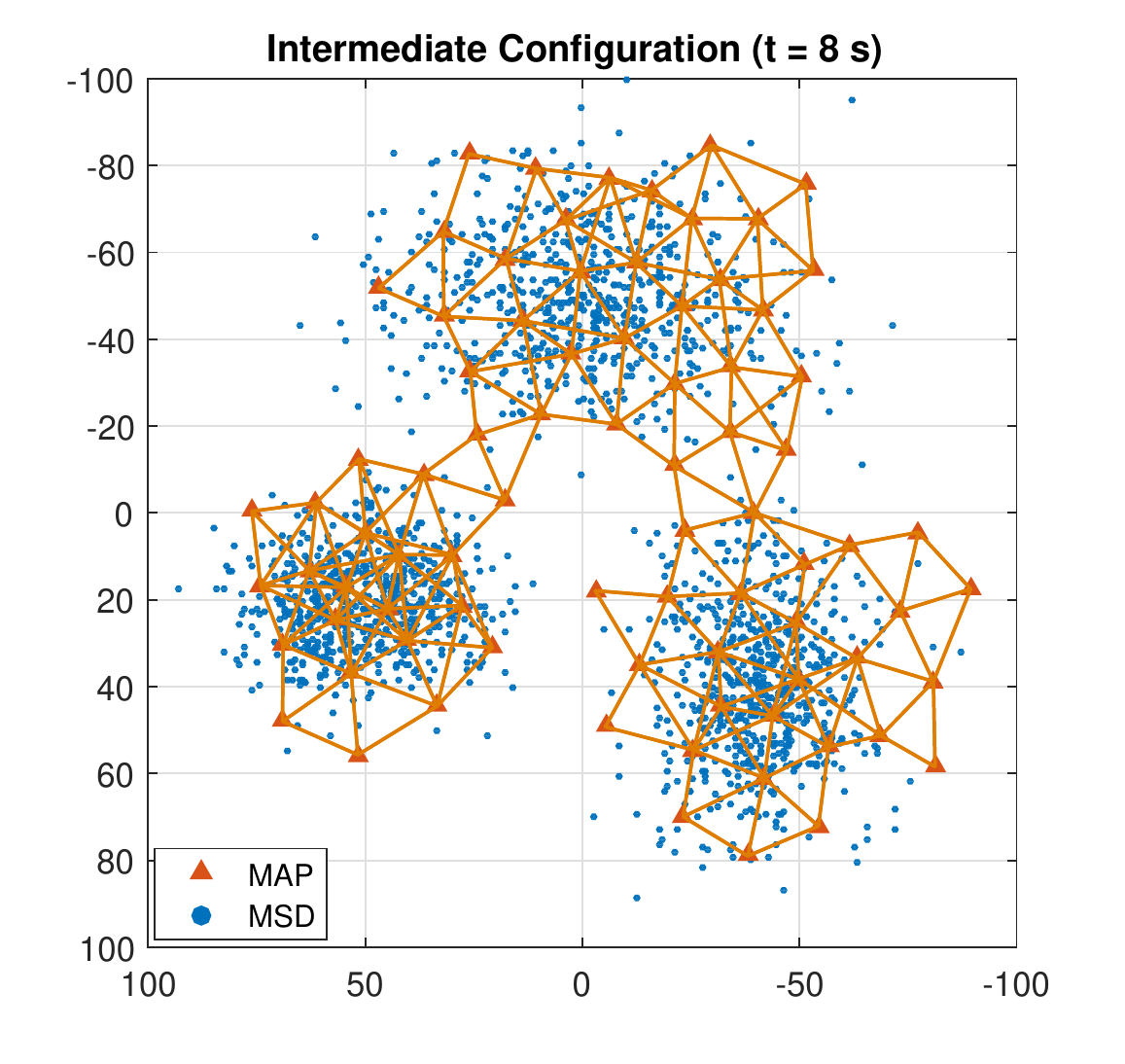} \label{config_t8}}\\
\subfloat[]{\includegraphics[width = 2.2in]{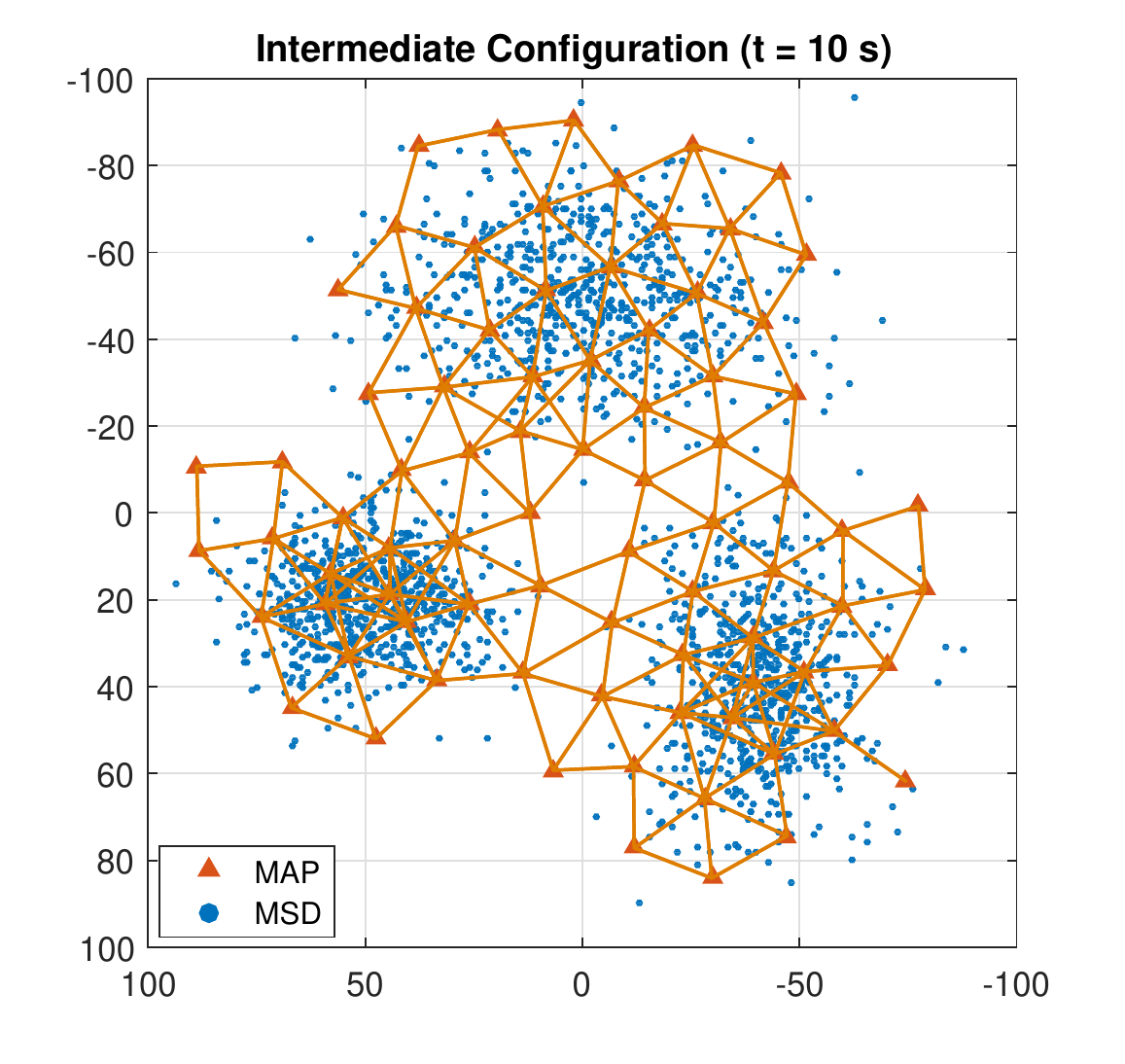} \label{config_t10}}
\subfloat[]{\includegraphics[width = 2.2in]{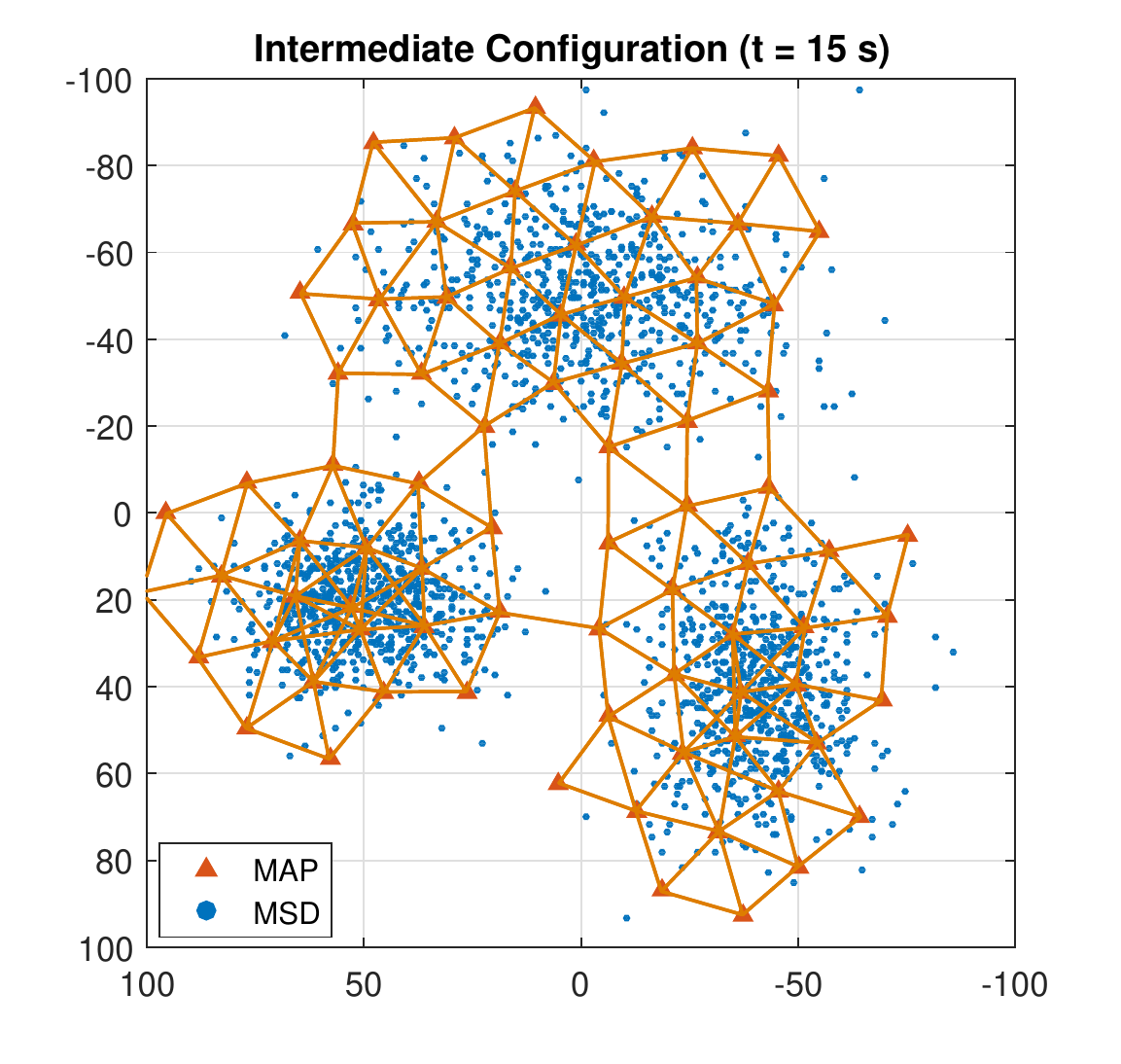} \label{config_t15}}
\subfloat[]{\includegraphics[width = 2.2in]{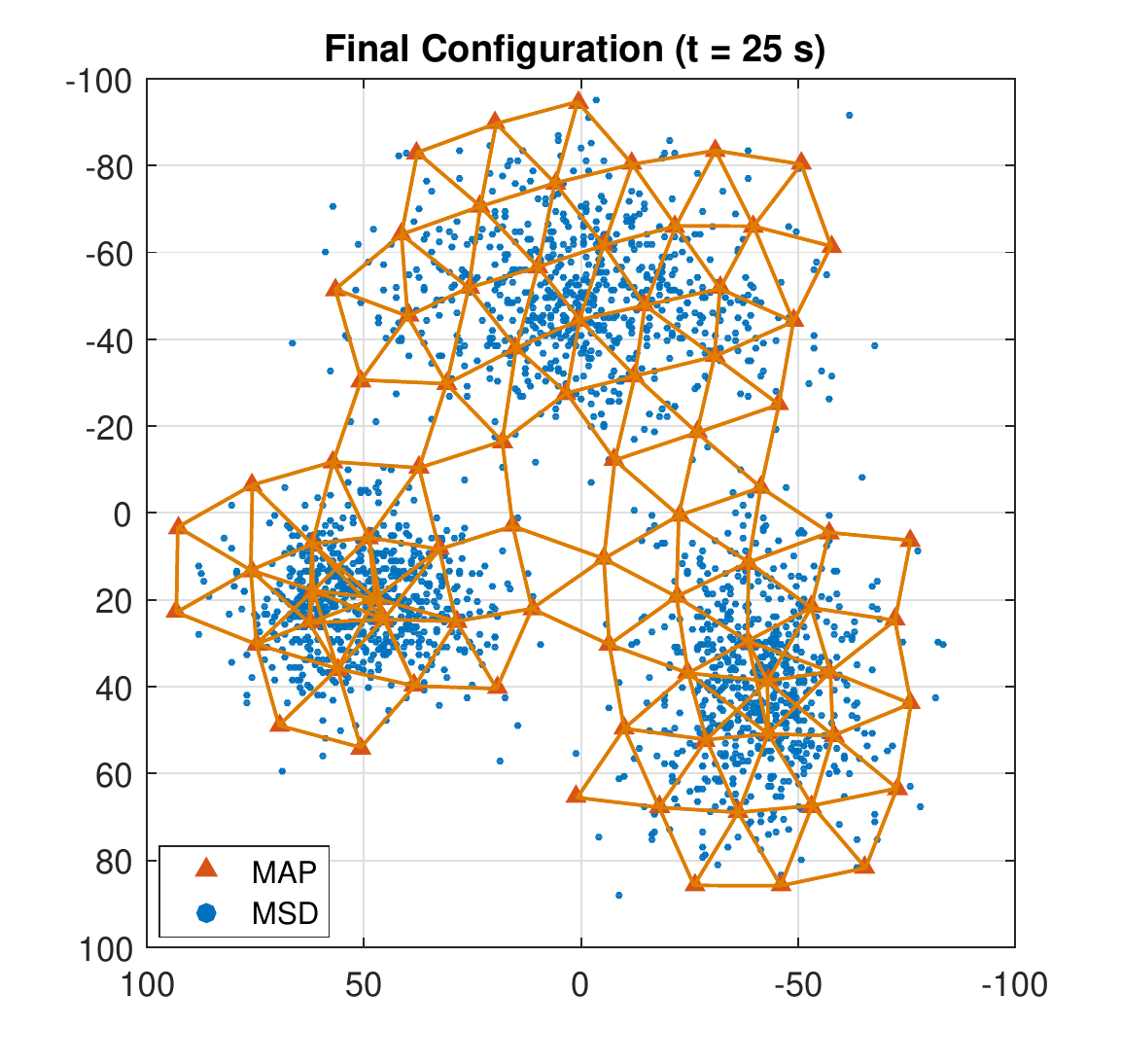} \label{config_t25}}
\caption{Example run of the cognitive algorithm showing snapshots of the MAP configuration at the initial stage, intermediate stages, and final stage after convergence. }
\label{Fig:convergence}
\end{figure*}
\begin{figure*}[t!]
\centering
\subfloat[]{\includegraphics[width = 2.2in]{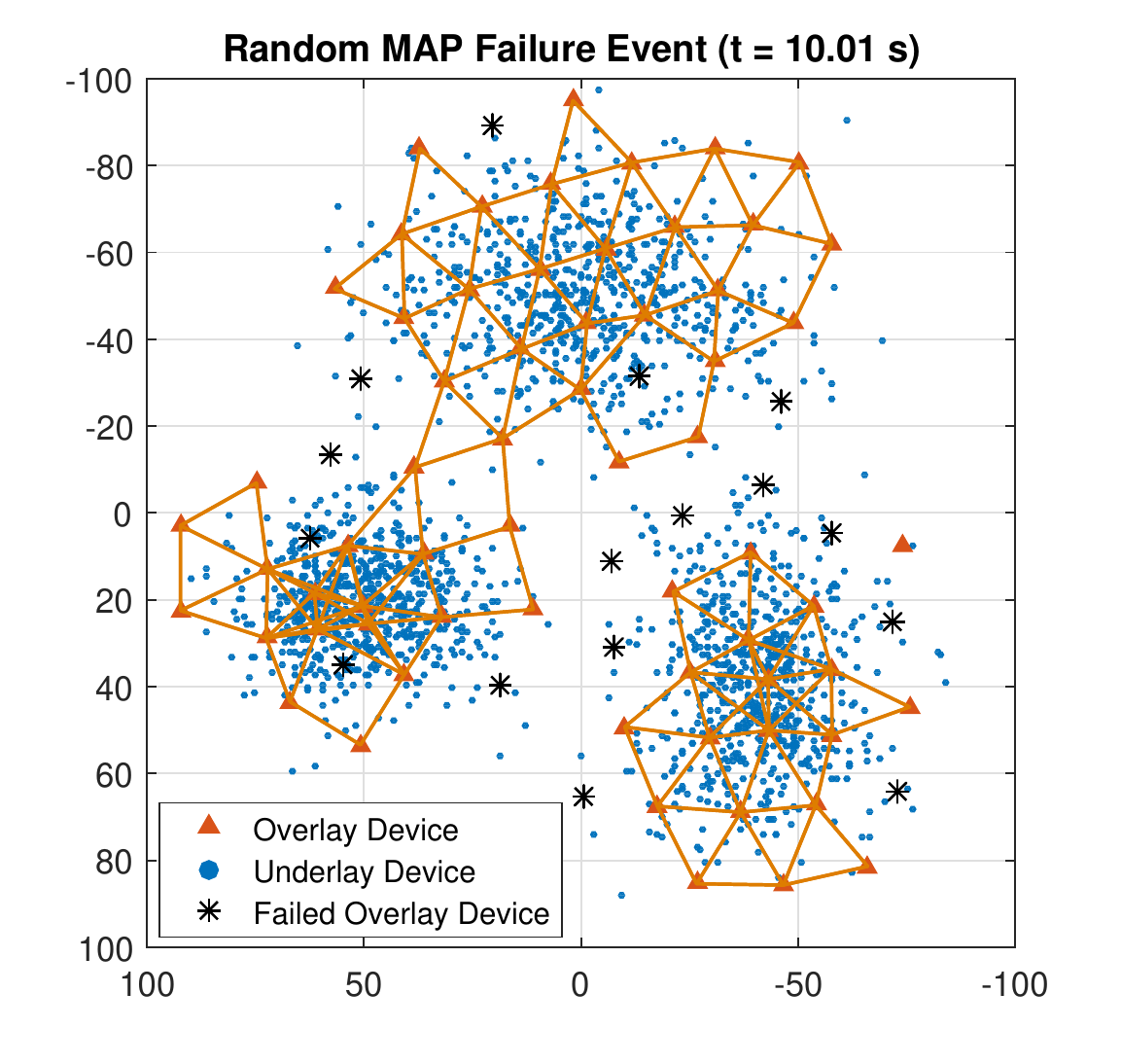} \label{recovery_t0}}
\subfloat[]{\includegraphics[width = 2.2in]{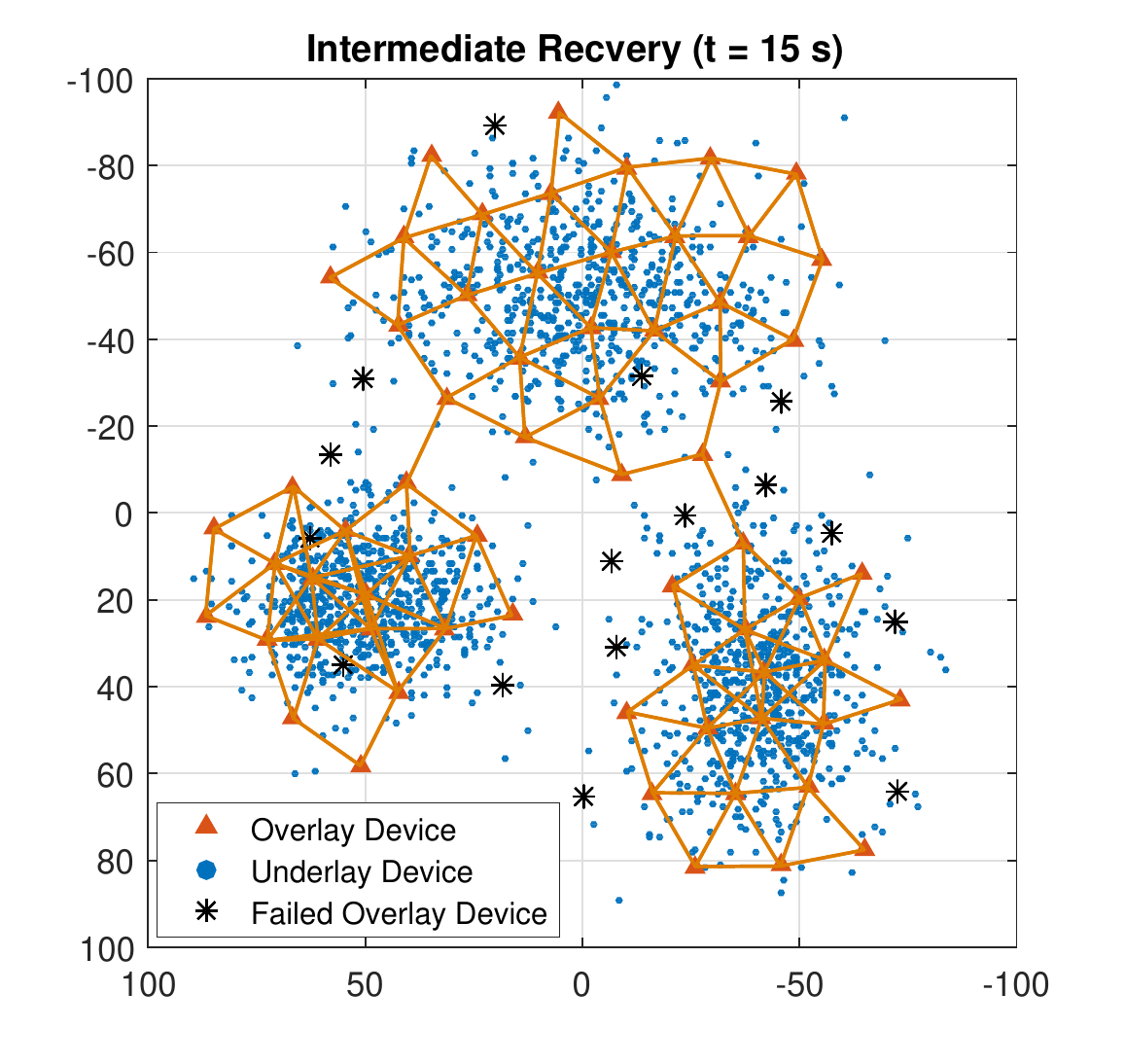} \label{recovery_t1}}
\subfloat[]{\includegraphics[width = 2.2in]{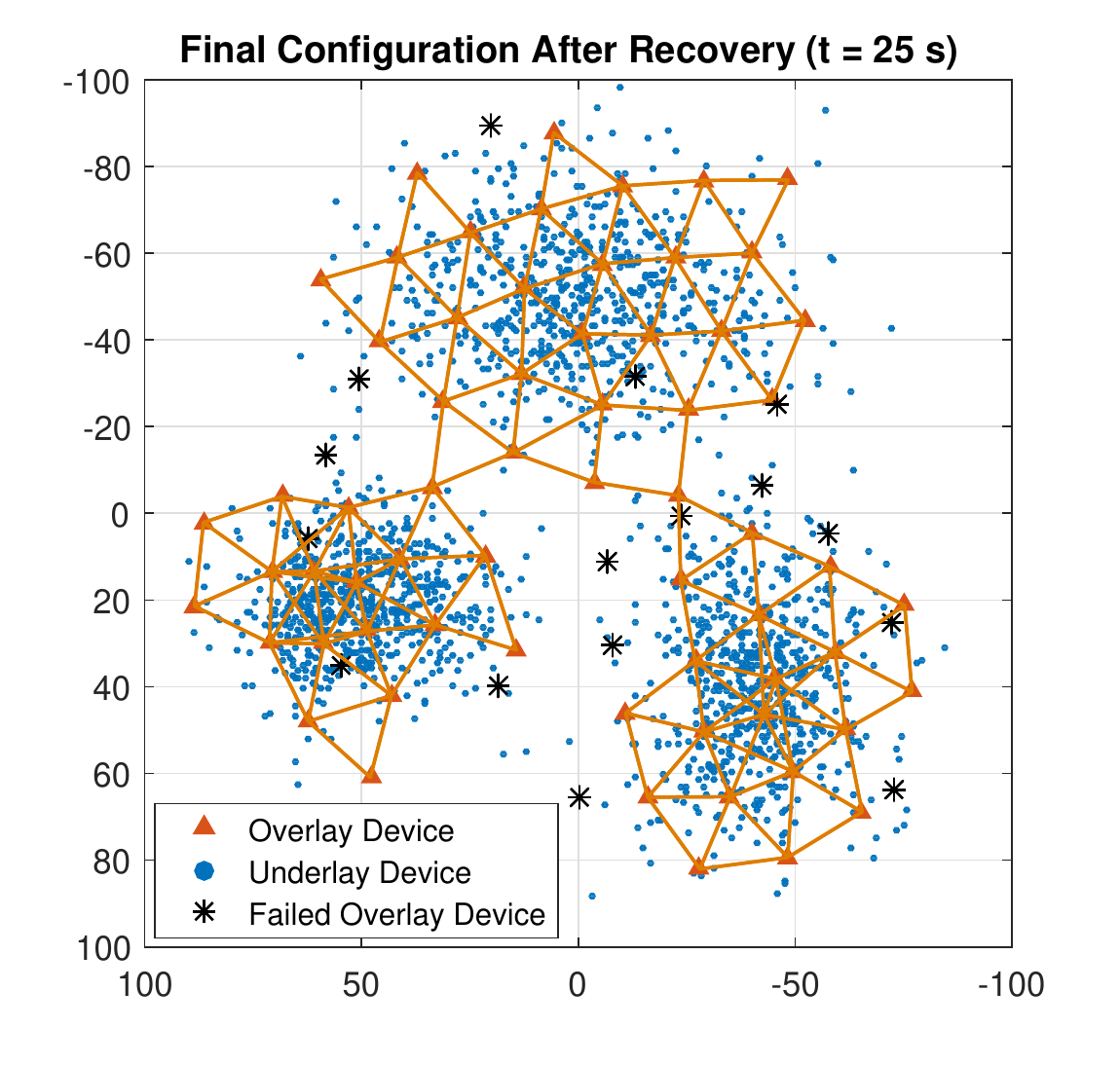} \label{recovery_t2}}
\caption{A random MAP failure event is induced at $t = 10$ s, making $20$\% of the MAPs unavailable. The cognitive framework adaptively re-configures itself to improve network connectivity as shown in the snapshots.}
\label{Fig:recovery}
\end{figure*}

%\begin{figure*}[t!]
%\centering
%\subfloat[]{\includegraphics[width = 2.4in]{Figures/Initial_config_1.eps} \label{initial_config}}
%\subfloat[]{\includegraphics[width = 2.4in]{Figures/Interim_config_1.eps} \label{interim_config}}
%\subfloat[]{\includegraphics[width = 2.4in]{Figures/Final_config_1.eps} \label{final_config}}\\
%\subfloat[]{\includegraphics[width = 2.4in]{Figures/Failure_config_1.eps} \label{failure_config}}
%\subfloat[]{\includegraphics[width = 2.4in]{Figures/Interim_recovery.eps} \label{interim_recovery}}
%\subfloat[]{\includegraphics[width = 2.4in]{Figures/Final_recovery_1.eps} \label{recovery_config}}
%\caption{Example run of the cognitive algorithm showing the initial intermediate and final configuration after convergence. A random MAP failure event occurs at $t = 12$ s, making $20$\% of the MAPs unavailable. The cognitive framework adaptively re-configures itself to improve network connectivity.}
%\label{results}
%\end{figure*}

\vspace{-0.0in}
\section{Results \& Discussion} \label{Sec:Results}

\subsection{Simulation Parameters}
In this section, we first describe the simulation settings before providing results on the performance of our proposed cognitive connectivity framework. Note that the selection of simulation parameters is made for illustrative purposes. We assume a bi-layer communication network comprising of MSDs such as IoT devices and MAPs such as UAVs. A set of $M = 2000$ MSDs are distributed in $\mathbb{R}^2$ according to a 2-D Gaussian mixture model with equal mixing proportions. The mean vectors are selected as $\mu_1 = [50,20]^T, \mu_2 = [0,-50]^T$, and $\mu_3 = [-40,40]^T$, and the covariance matrices are selected as follows:
\begin{align}
\hspace{-0.3cm}\Sigma_1 = \hspace{-0.1cm}
\arraycolsep=1.5pt\def\arraystretch{1.4}
  \left[ {\begin{array}{cc}
   200 \quad & 0 \\
   0 \quad & 100 \\
  \end{array} } \right],
\Sigma_2 = \hspace{-0.1cm}
  \left[ {\begin{array}{cc}
   500 \quad & 0 \\
   0 \quad & 200 \\
  \end{array} } \right],
  \Sigma_3 = \hspace{-0.1cm}
  \left[ {\begin{array}{cc}
   150 \quad & 0 \\
   0 \quad & 300 \\
  \end{array} } \right]\hspace{-0.1cm}.
\end{align}

The mobility of the MSDs is modeled by a scaled uniform random noise at each time step, i.e., $y_i(k+1) = y_i(k) + s\xi$, where $\xi \sim$ Uniform($[-1,1] \times [-1,1]$), where $s = 0.2$ represents the scale. The $L = 80$ MAPs are initially distributed uniformly in the plane perpendicular to the vector $[0,0,h]^T$ where the altitude $h$ is selected as 20 m. The initial velocity vectors of the MAPs are selected uniformly at random from the box $[-1,-2]^2$. \textcolor{black}{We assume that the utility function used to define the quality of the communication link between the MSDs and the MAPs is $\Phi(i,j) = \kappa \|y_i - q_j \|^{-\eta}$, where $\kappa \in \mathbb{R}^+$ is the transmission power and $\eta\in \mathbb{R}^+$ represents the path-loss exponent. Furthermore, we assume that $\kappa = 1$ and $\eta = 4$.}
A list of all the remaining parameter values used during the simulations is provided in Table~\ref{parameters}. We run the simulations with a step size of $\Delta = T_s$ up to $t = 25$ s.

\textcolor{black}{For the purpose of simulation, the strength of communication links between the MAPs and their neighbors is characterized by the cutoff function in (3) with thresholds $z_1 = \gamma = 0.2$ and $z_0 = 1$. In other words, we have abstracted the wireless propagation details by using the cutoff function $\alpha_{\{\gamma, 1\}}$ which is parameterized by $\gamma$. Hence, each MAP only requires the knowledge of the distance to its neighbouring MAPs that are inside its communication range and the threshold $\gamma$ to evaluate its control strategy using (8). The particular choice of $\gamma$ implies that two MAPs are considered to reliably communicate if the normalized distance between them is within $20$\% of the maximum normalized communication range. On the other hand, two MAPs are considered completely unable to communicate if the normalized distance is greater than the maximum normalized communication range. However, for practical implementation, the threshold $\gamma$ needs to be accurately determined based on the radio propagation model used and the communication reliability requirements.
}

In Fig.~\ref{Fig:convergence}, we provide an example run of the proposed cognitive connectivity framework. We show a top view of the network for the sake of clear presentation. Fig.~\ref{config_t0} shows the initial configuration at $t = 0$, when the MAPs are randomly deployed over an underlying population of MSDs. The MSDs keep moving with time in the goal directions as observed by the different MSD locations in~\cref{config_t4_8,config_t8,config_t10,config_t15}. As the proposed cognitive connectivity framework evolves, the MAPs tend to move toward the closest group of MSDs as shown in Fig.~\ref{config_t4_8}. Finally, when the framework converges, the MAPs develop a desirable connected formation hovering over the MSDs as shown by Fig.~\ref{config_t25}. It should be noted that the MAPs are located closer to each other in areas where MSDs are densely deployed, such as around cluster centers. In areas where the MSDs are sparsely located, the MAPs develop a regular formation.

%Fig. is useful as it shows that at higher failure proportion, the MAP network becomes disconnected.

\begin{figure}[t]
  \centering
  \includegraphics[width=3in]{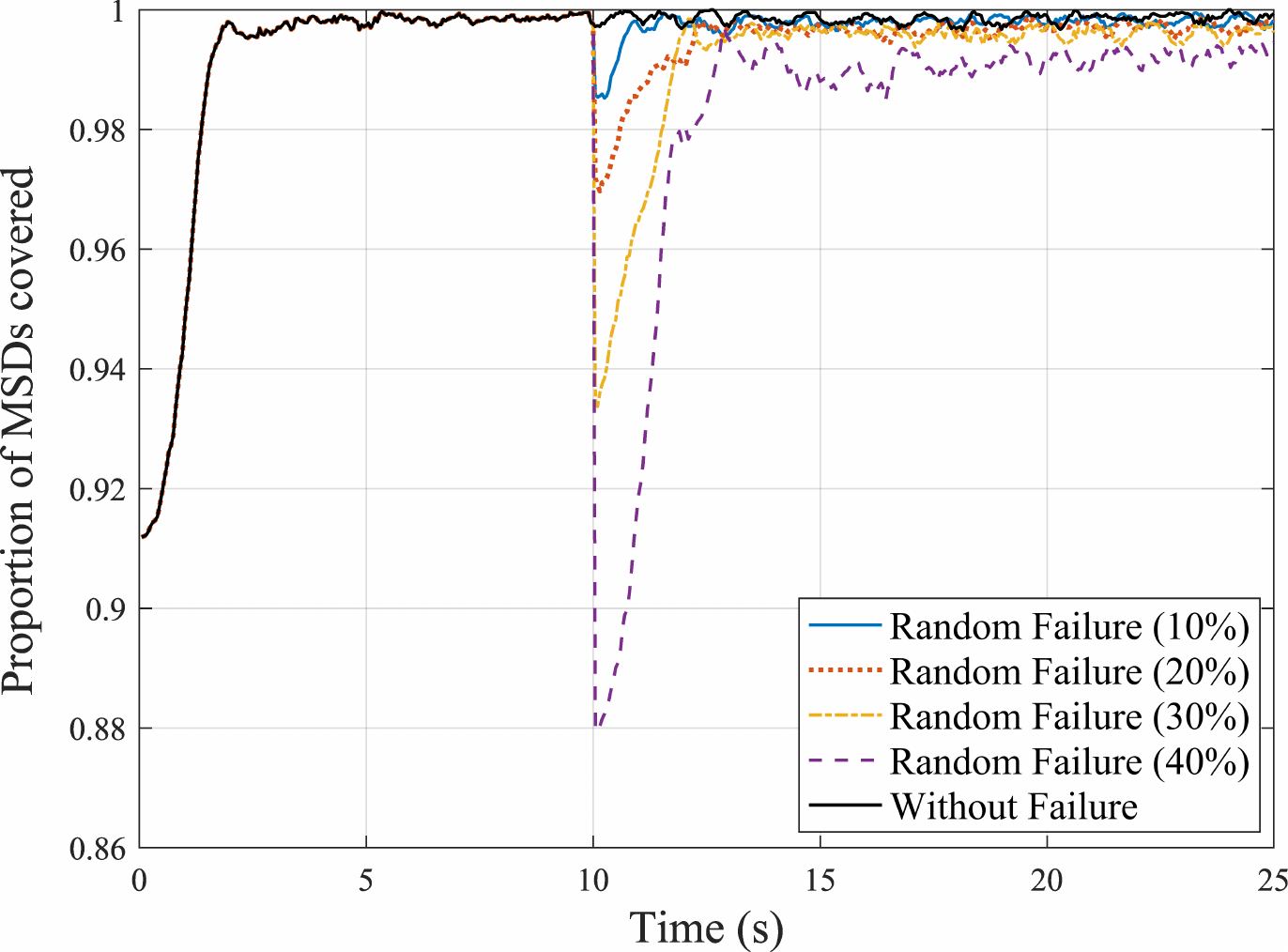}\\
  \caption{Proportion of MSDs covered by the MAPs.}\label{pc_covered}
\end{figure}

\begin{figure}[t]
  \centering
  \includegraphics[width=3in]{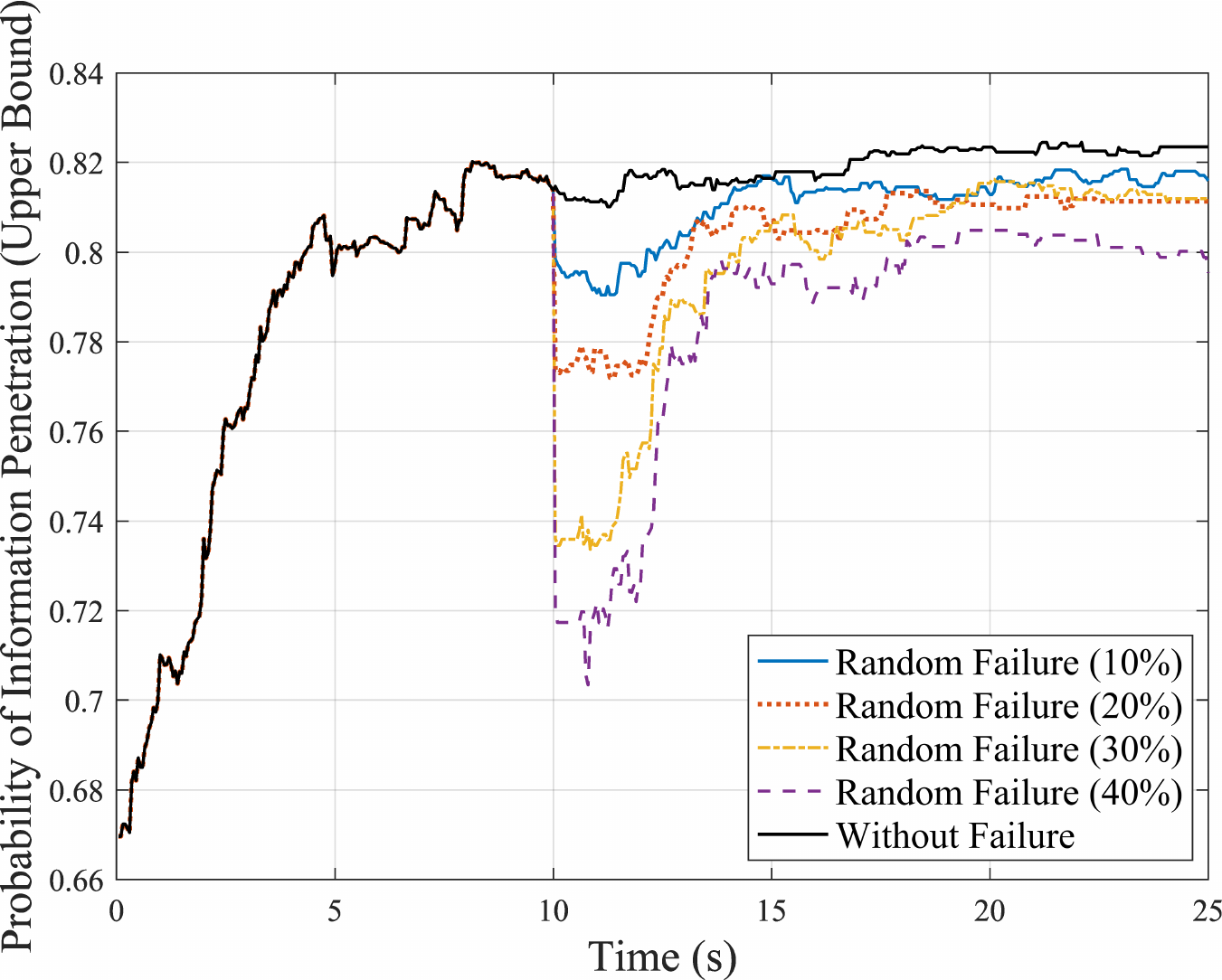}\\
  \caption{Probability of information penetration in the D2D enabled MAP network.}\label{pc_informed}
\end{figure}

\begin{figure}[t]
  \centering
  \includegraphics[width=3in]{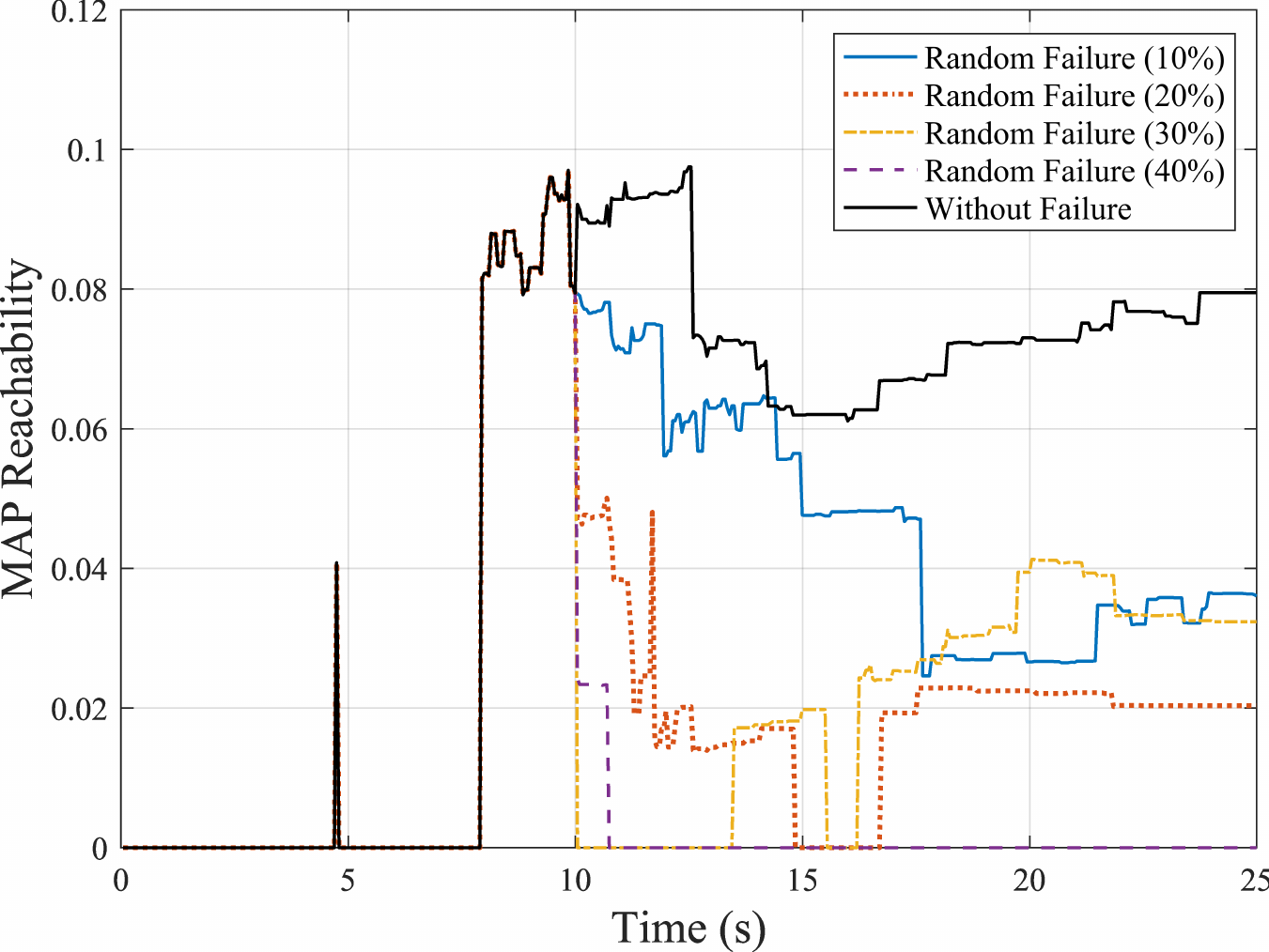}\\
  \caption{Reachability of MAPs determined by the algebraic connectivity.}\label{algebraic}
\end{figure}

\begin{figure*}[t!]
\centering
\subfloat[Circle Packing Approach]{\includegraphics[width = 2.2in]{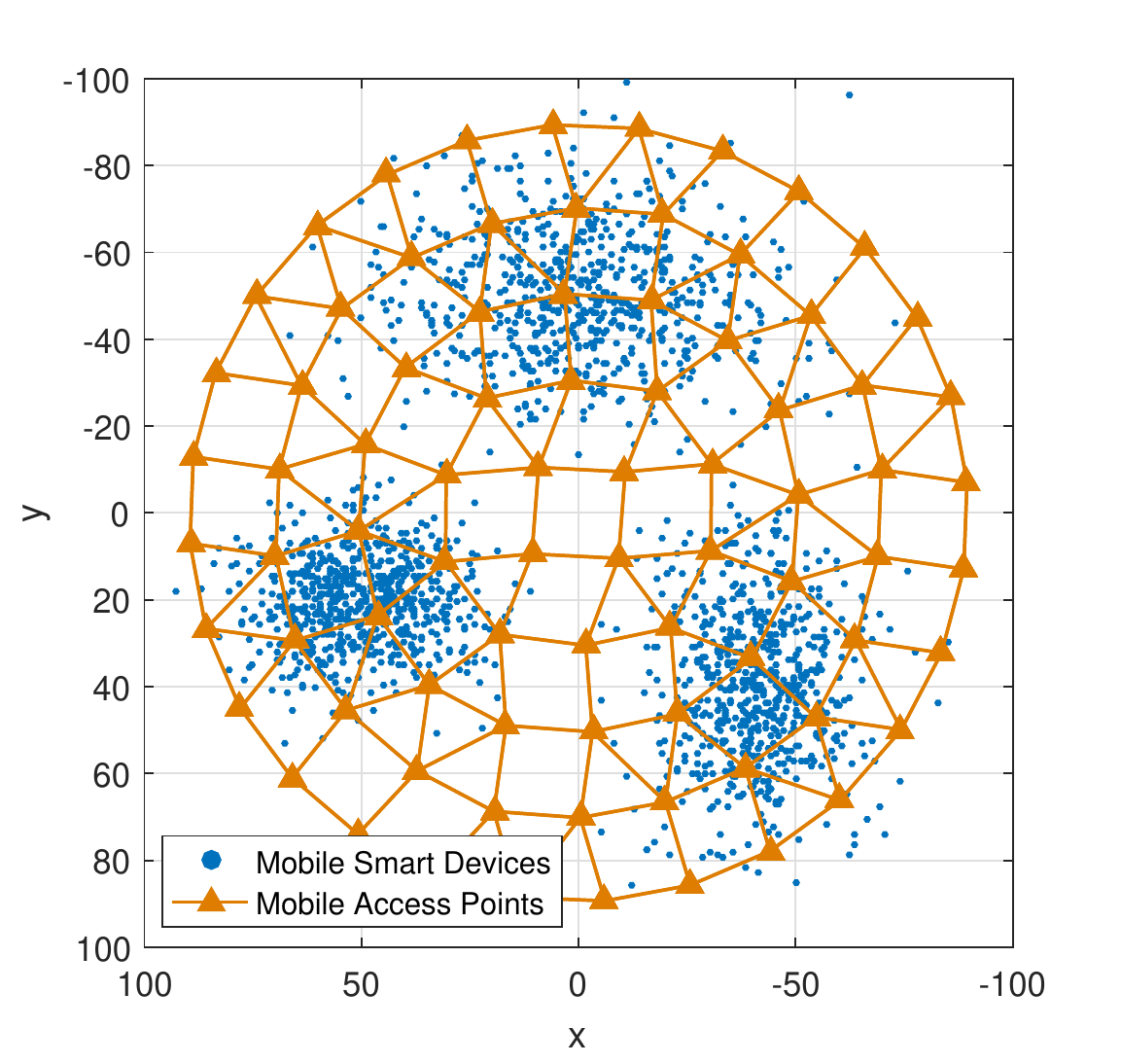} \label{Fig:circle_packing}}
\subfloat[p-median Approach]{\includegraphics[width = 2.2in]{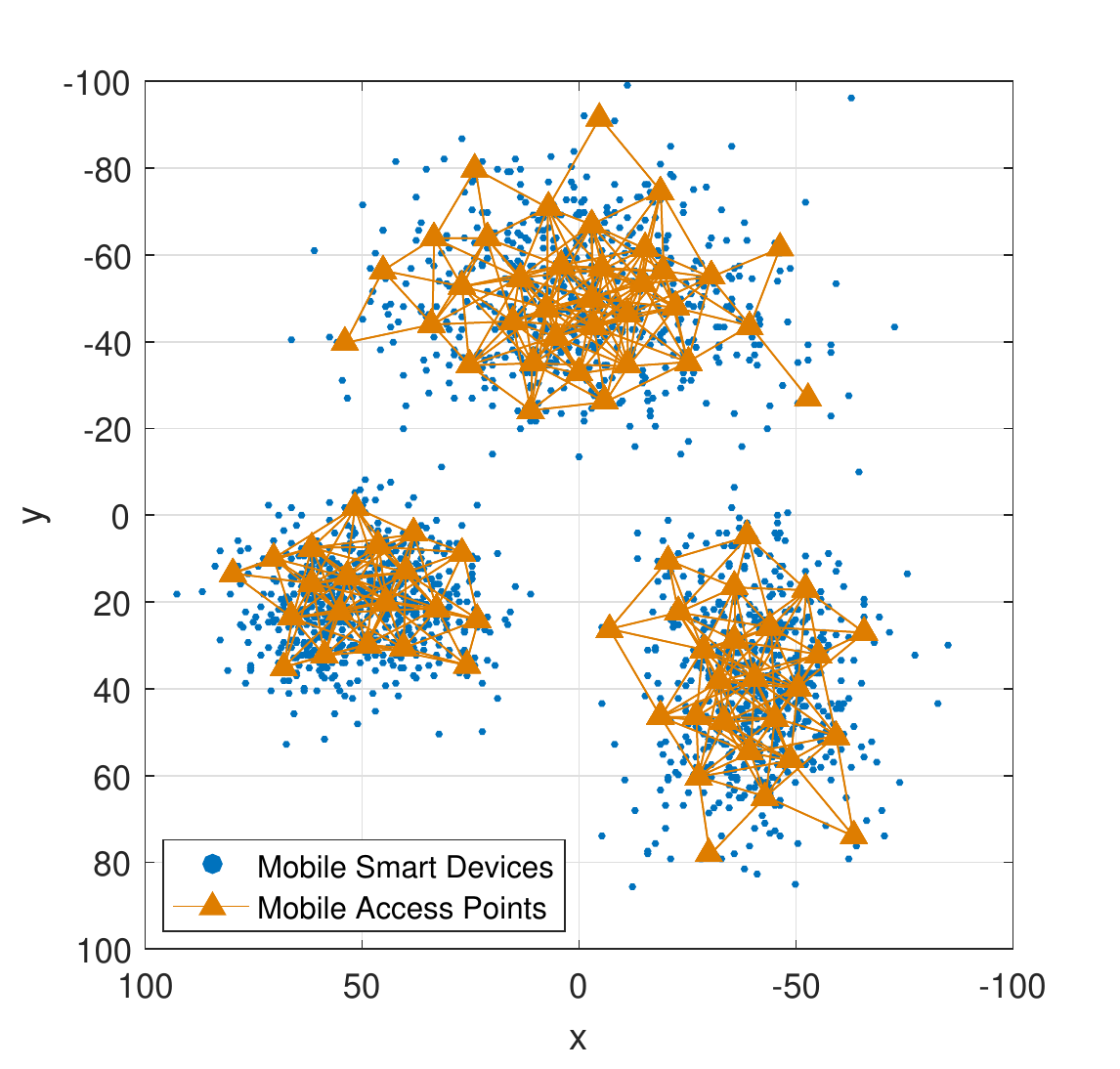} \label{Fig:p_median}}
\subfloat[Proposed Dynamic Approach]{\includegraphics[width = 2.2in]{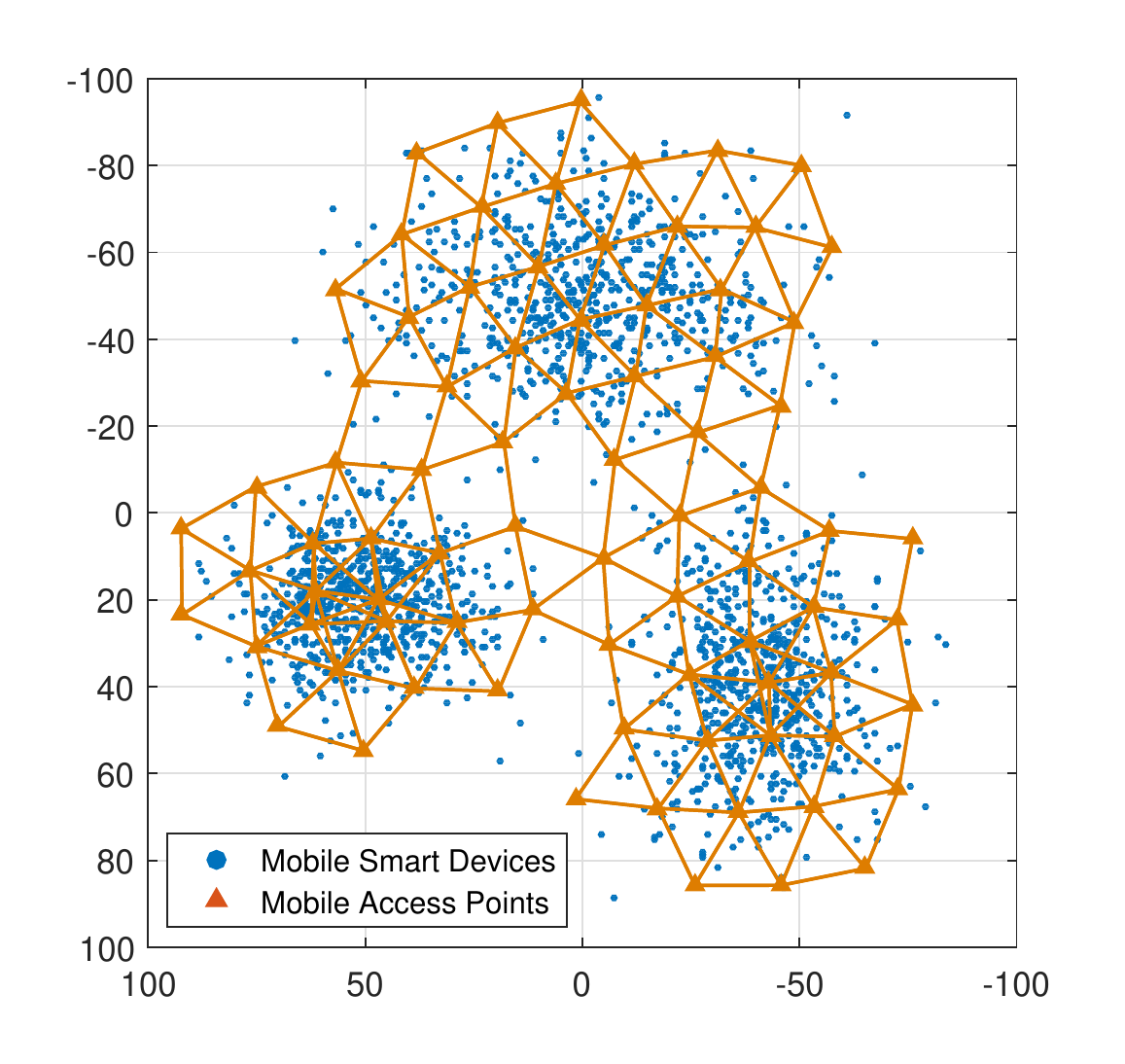} \label{Fig:Proposed}}
\caption{Top view of final configurations of MAPs in comparison to the proposed approach.}
\label{results}
\end{figure*}
\subsection{Resilience}
In this section, we investigate the impact of a random MAP failure event on the connectivity of the network and evaluate the response of the proposed cognitive framework in such a situation. Fig.~\ref{Fig:recovery} shows an induced random failure of 20\% of the MAPs, which results in loss of coverage to the MSDs as well as reduction in the connectivity of the MAPs. The proposed framework immediately starts responding to the coverage gap created by the MAP failure and tends to reconfigure itself as shown by the intermediate snapshot in Fig.~\ref{recovery_t1}. Eventually, the framework converges leading to a coverage maximizing configuration while maintaining connectivity of the MAPs as shown in Fig.~\ref{recovery_t2}. To test the resilience of the proposed framework under different levels of MAP failure events, we simulate varying severity of device failure events from 10\% to 40\% failed MAPs in the overlay network. Fig.~\ref{pc_covered} illustrates the proportion of MSDs that are covered by the MAPs. Without any failure, it is observed that the proposed framework successively improves the coverage until almost all the MSDs are covered. Notice that the coverage fluctuations occur due to the continuously mobile MSDs. Once the device failure event occurs at around $t = 10$ s, the framework responds and is able to quickly restore maximum coverage except in the case of 40\% failure, in which the coverage is not fully restored. In Fig.~\ref{pc_informed}, we plot the probability of information penetration in the MAP network. It can be observed that the framework is able to recover up to 97\% of the original value. However, it is important to note that the probability of information dissemination is an upper bound and does not provide information about the reachability of the MAPs. In this situation, Fig.~\ref{algebraic}, which shows the algebraic connectivity of the MAPs, proves to be extremely useful. It is observed that when the failure proportion is 10\%, 20\%, or 30\%, the reachability can be restored by the proposed framework, as indicated by the nonzero algebraic connectivity. However, in the event of 40\% failure, the algebraic connectivity remains zero even after reconfiguration, which implies that the MAP network is no more connected. However, since the probability of information penetration is still high, it implies that the MAPs have also clustered around the MSD clusters thus providing effective intra-cluster connectivity.
Note that there is a spike in the MAP reachability at $t = 4.8$ s. This is because the dynamic MAPs momentarily become connected at $t = 4.8$ s as shown by Fig.~\ref{recovery_t1} before being disconnected by a single link.

Furthermore, the framework is also resilient to many other variations in the network. For instance, in the case of crowd mobility, the proposed cognitive framework can readily adapt to the changing positions of the users without additional complexity. However, existing approaches in literature cannot adapt to such changes in realtime. This is because they require global network information along with the need to solve the NP-hard optimization problem repeatedly. Similarly, the proposed framework is also resilient to a wide variety of cyber-physical attacks as it uses a cognitive feedback loop to constantly update the states of each MAP independently of the others which makes it less vulnerable to cyber threats.

\begin{figure}[t]
  \centering
  % Requires \usepackage{graphicx}
  \includegraphics[width=3in]{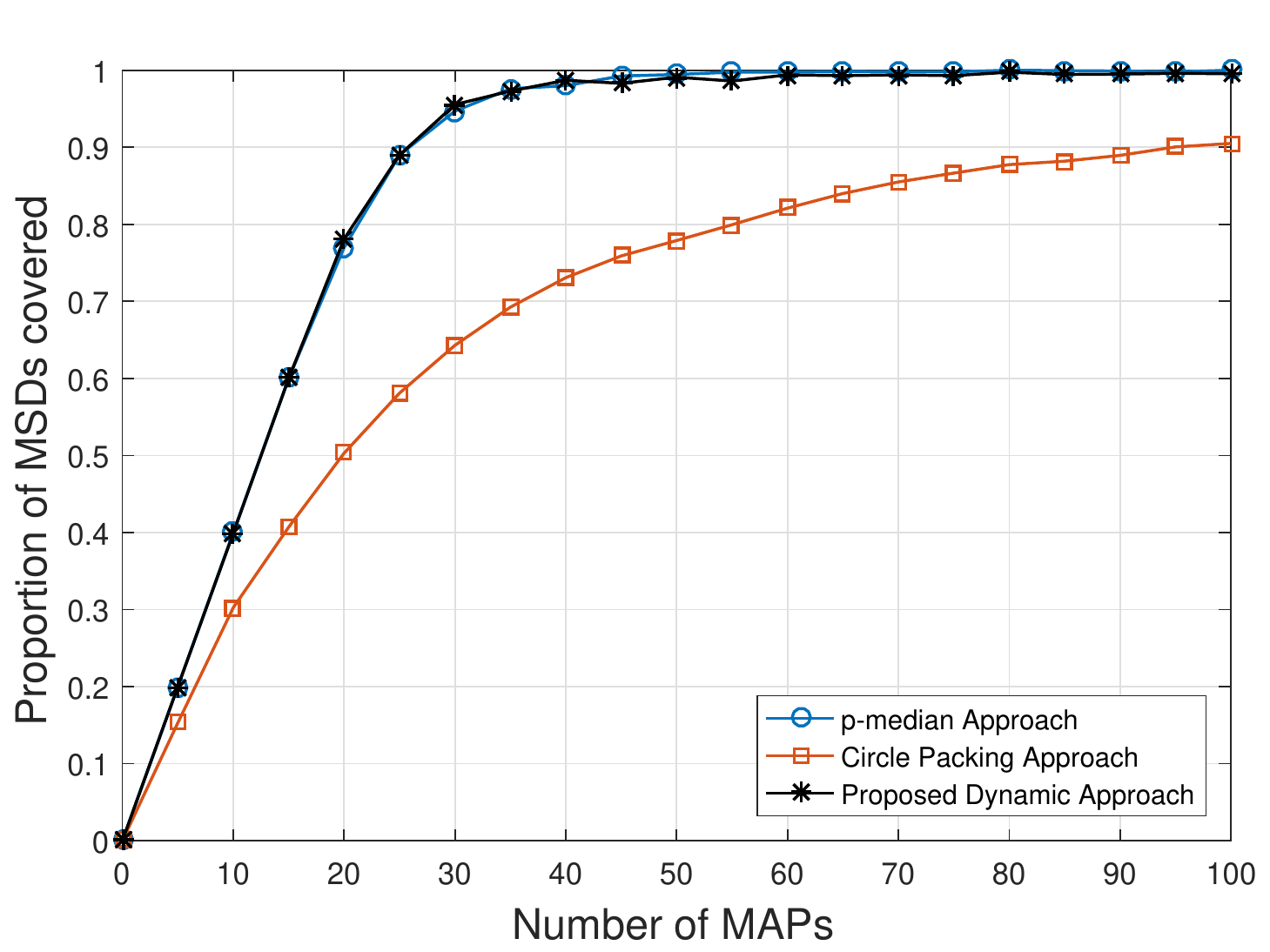}\\
  \caption{Comparison of coverage of MSDs with baseline algorithms.}\label{Fig:coverage_comparison}
\end{figure}

\begin{figure}[t]
  \centering
  % Requires \usepackage{graphicx}
  \includegraphics[width=3in]{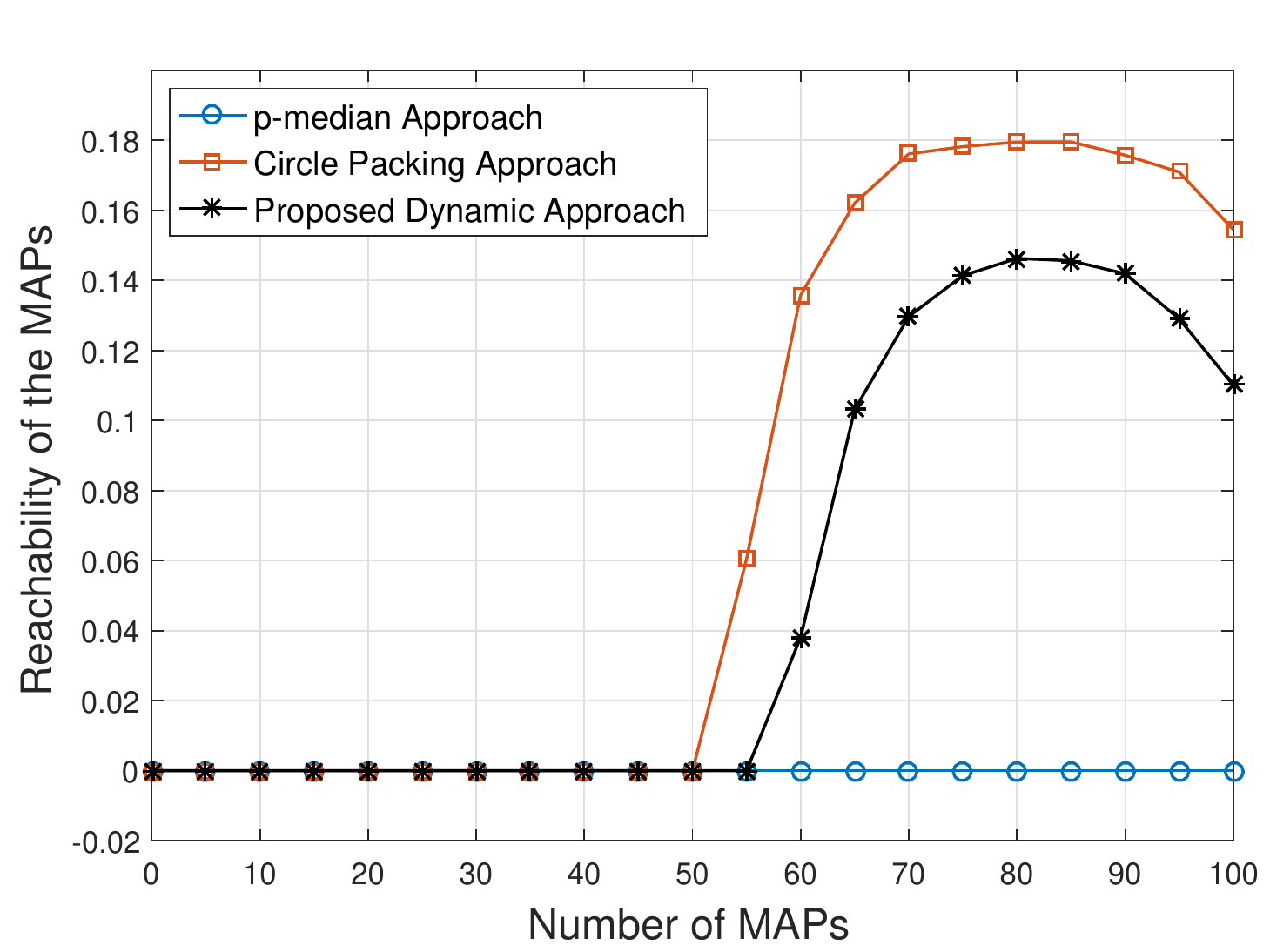}\\
  \caption{Comparison of reachability of MAPs with baseline algorithms.}\label{Fig:reachability_comparison}
\end{figure}

\begin{figure}[t]
  \centering
  % Requires \usepackage{graphicx}
  \includegraphics[width=3in]{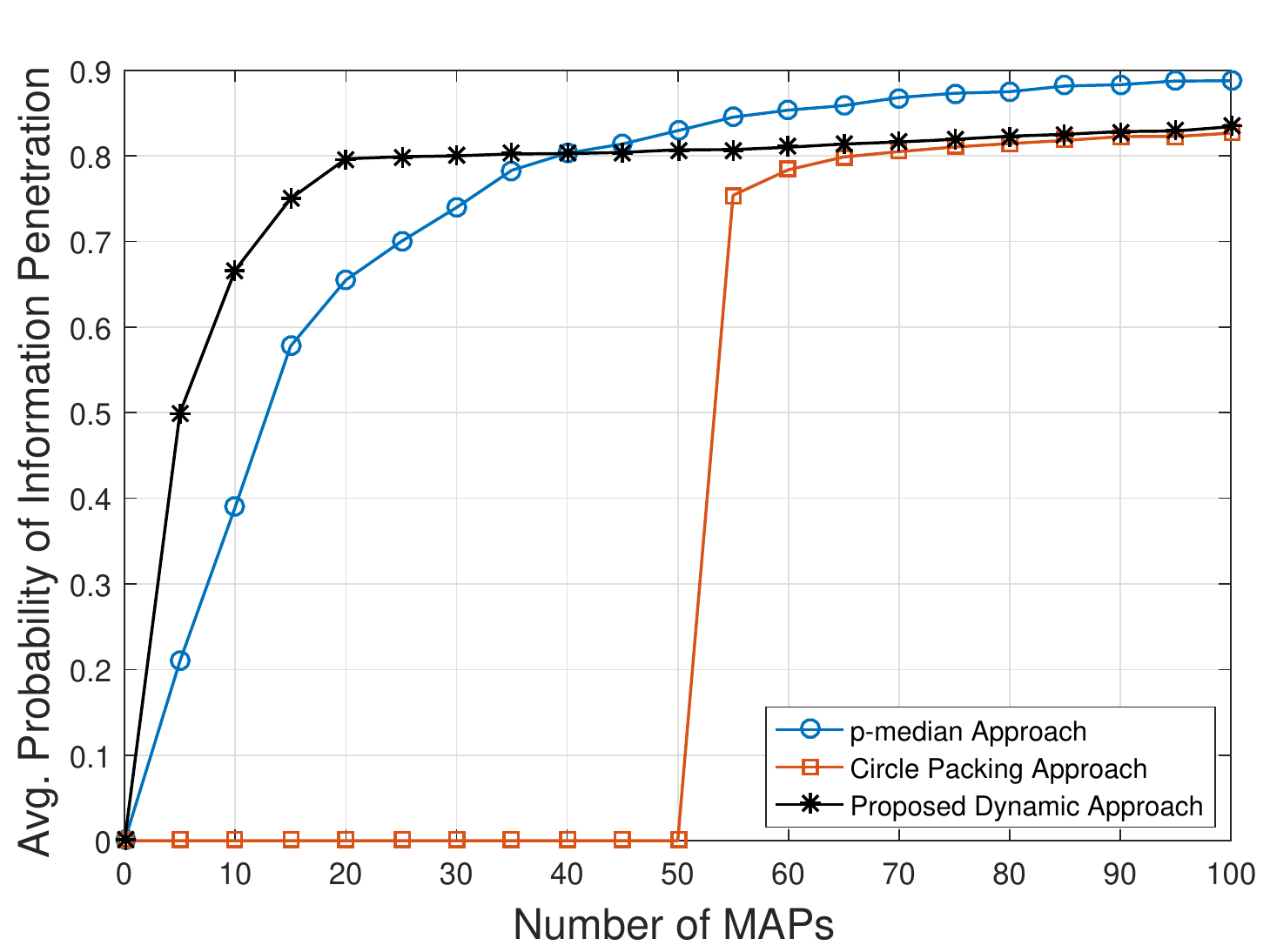}\\
  \caption{Comparison of probability of information dissemination in the MAPs with baseline algorithms.}\label{Fig:dissemination_comparison}
\end{figure}

\subsection{Comparison}
In this section, we compare the performance of our proposed algorithm with existing frameworks. We use the two widely used approaches in literature, i.e., the p-median approach and the circle packing approach as baseline algorithms. Since the compared solution algorithms provide centralized solutions, we ignore the effect of mobility of the MSDs in this section for fair comparison. Before measuring the performance in terms of the metrics, we illustrate an example MAP configuration obtained using each of the approaches in Fig.~\ref{results}.
Fig.~\ref{Fig:circle_packing} shows the optimal configuration of MAPs obtained using the circle packing solution. It can be observed that the MAPs are tightly packed inside a circular region covering the MSDs.
Fig.~\ref{Fig:p_median} shows the optimal configuration achieved using the p-median solution. As we would expect, all the MAPs are concentrated around the MSDs and since the MSDs are distributed in spatial clusters, the MAPs are unable to maintain connectivity between them. Finally, Fig.~\ref{Fig:Proposed} shows the MAP configuration achieved after convergence of our proposed cognitive framework. Note that our proposed framework leads to a configuration that covers all the MSDs and also maintains connectivity between the MAPs.

Next, we compare the performance of the proposed algorithm with the baseline in terms of the metrics set for performance evaluation. Fig.~\ref{Fig:coverage_comparison} shows the proportion of MSDs covered by the MAPs for each of the algorithms. It can be observed that the proposed dynamic algorithm and the p-median approach result lead to complete coverage of the MSDs with significantly less number of MAPs ($\sim$20) as compared to the circle packing approach ($>$100). The primary reason for this is the restriction of MAPs to fill a circle as packing circles inside an arbitrary region is NP-hard. Moreover, since the MAPs cannot be positioned arbitrarily close to each other, so there is loss of coverages when the number of users under the influence of a MAP exceeds its capacity. In Fig.~\ref{Fig:reachability_comparison}, we plot the reachability or algebraic connectivity of the MAPs using the three algorithms. It can be observed that the p-median approach does not lead to a connected network even with very high number of MAPs. On the other hand, both the proposed algorithm and the circle packing approach leads to a connected configuration, i.e., the Fiedler value is non-zero, if the number of MAPs is sufficiently high ($>$40). Note that the magnitude of the Fiedler value reflects the tendency of the network to become disconnected. Since the circle packing approach leads to a well connected configuration, it is harder to make the network disconnected, which is why it has a higher Fiedler value. Finally, Fig.~\ref{Fig:dissemination_comparison} shows the average probability of information dissemination in the MAP network. It can be observed that the proposed cognitive approach results in a high average information penetration probability with a small number of MAPs ($\sim20$). On the other hand, the other approached require significantly higher number of MAPs to achieve a similar level of average information dissemination. In light of the above comparisons, it is clear that the proposed cognitive framework significantly outperforms other frameworks available in literature. It is also pertinent to mention that the baseline algorithms are not dynamic and hence would perform even \textcolor{black}{worse} under the mobility of MSDs.

\vspace{-0.0in}
\section{Conclusion} \label{Sec:Conclusion}
In this paper, we present a cognitive connectivity framework that is able to reconfigure itself autonomically in a distributed manner to interconnect spatially dispersed smart devices thus enabling the Internet of things in remote environments.
Resilience of connectivity has been investigated in response to the mobility of the underlay network as well as random device failures in the overlay network.
It is shown that if sufficient number of overlay devices are available, then the developed distributed framework leads to high network connectivity which is resilient to mobility and device failures. However, if sufficient overlay devices are not deployed, the framework tends to provide connectivity locally to the devices in each cluster of the underlay network. A comparison of the proposed approach with existing approaches for placement of BSs reveals significant superiority in terms of the number of BSs required to achieve coverage and the overall connectivity of the devices.

We believe that this work provides a useful platform for the development of more sophisticated and efficient algorithms to achieve a variety of objectives in aerial communications using UAVs. Future directions in this work can investigate on ways to make the framework completely distributed. The local observations of MAPs can be used to form a consensus about the locations of the MSDs. Another possible direction to this line of research is to allow MAPs to operate in multiple modes to enable connectivity between a diverse pattern of locations of the MSDs.

\vspace{-0.1in}
\bibliographystyle{IEEEtran}
\bibliography{new_references}

\begin{IEEEbiography}
    [{\includegraphics[width=1in,height=1.25in,clip,keepaspectratio]{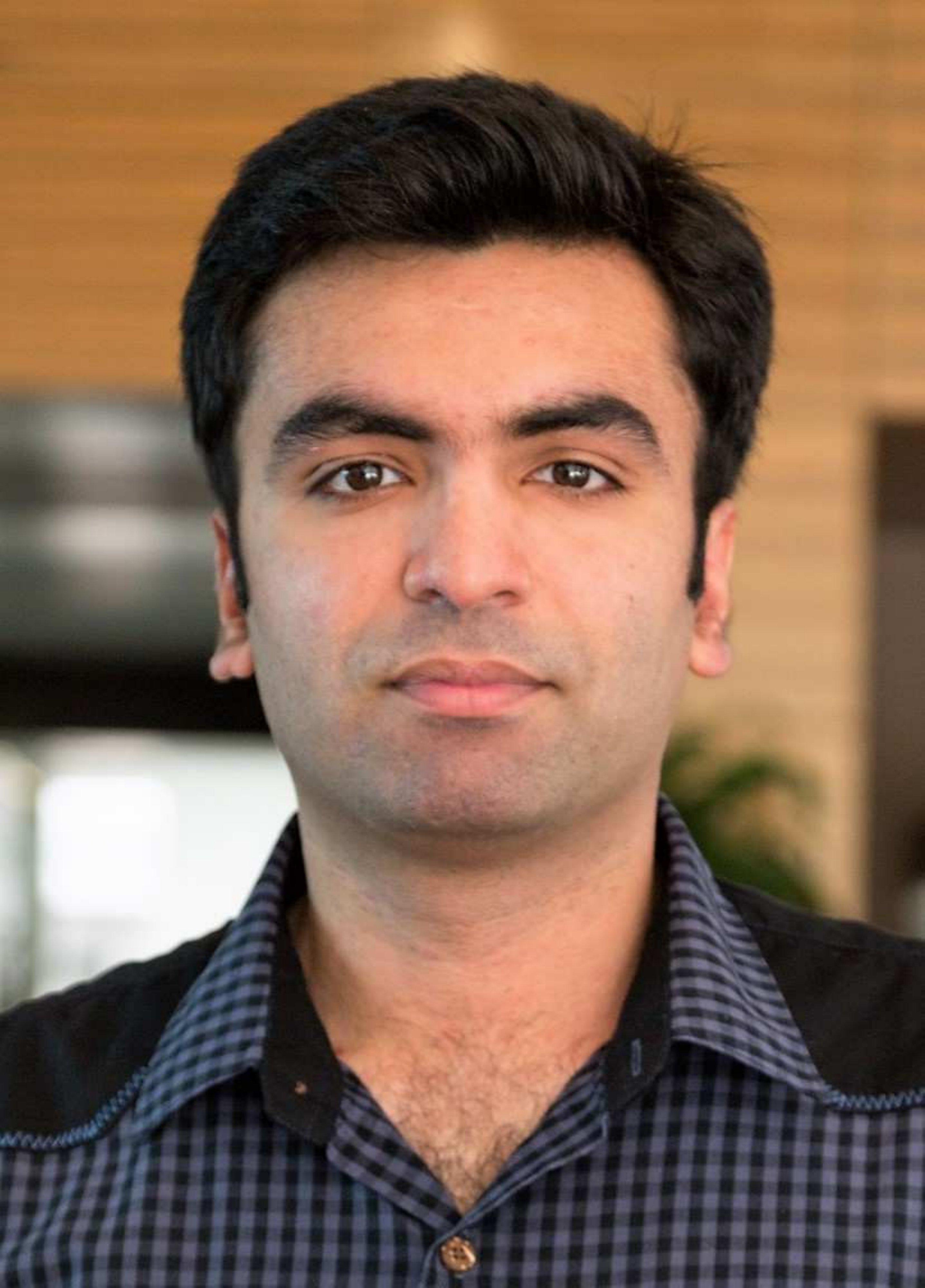}}]{Muhammed Junaid Farooq} received the B.S. degree in electrical engineering from the School of Electrical Engineering and Computer Science (SEECS), National University of Sciences and Technology (NUST), Islamabad, Pakistan, the M.S. degree in electrical engineering from the King Abdullah University of Science and Technology (KAUST), Thuwal, Saudi Arabia, in 2013 and 2015, respectively. Then, he was a Research Assistant with the Qatar Mobility Innovations Center (QMIC), Qatar Science and Technology Park (QSTP), Doha, Qatar. Currently, he is a PhD student at the Tandon School of Engineering, New York University (NYU), Brooklyn, New York. His research interests include modeling, analysis and optimization of wireless communication systems, cyber-physical systems, and the Internet of things. He is a recipient of the President's Gold Medal for academic excellence from NUST, the Ernst Weber Fellowship Award for graduate studies and the Athanasios Papoulis Award for graduate teaching excellence from the department of Electrical \& Computer Engineering (ECE) at NYU Tandon School of Engineering.
\end{IEEEbiography}

\begin{IEEEbiography}
    [{\includegraphics[width=1in,height=1.25in,clip,keepaspectratio]{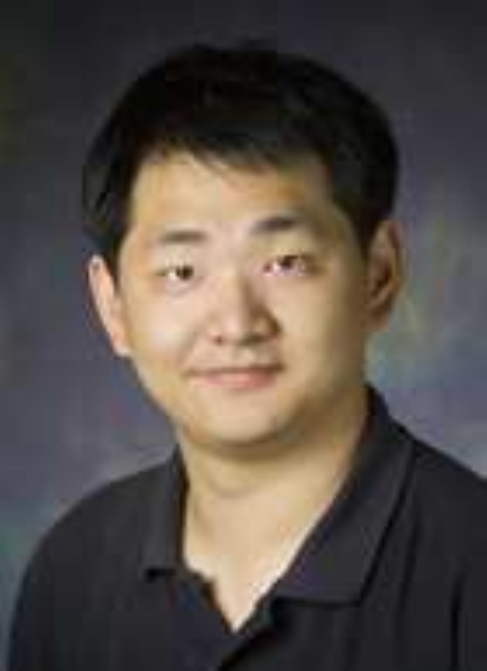}}]{Quanyan Zhu} (S'04, M'12) received B. Eng. in Honors Electrical Engineering from McGill University in 2006, M.A.Sc. from University of Toronto in 2008, and Ph.D. from the University of Illinois at Urbana-Champaign (UIUC) in 2013. After stints at Princeton University, he is currently an assistant professor at the Department of Electrical and Computer Engineering, New York University. He is a recipient of many awards including NSERC Canada Graduate Scholarship (CGS), Mavis Future Faculty Fellowships, and NSERC Postdoctoral Fellowship (PDF). He spearheaded and chaired INFOCOM Workshop on Communications and Control on Smart Energy Systems (CCSES), and Midwest Workshop on Control and Game Theory (WCGT). His current research interests include Internet of things, cyber-physical systems, security and privacy, and system and control.
\end{IEEEbiography}

% that's all folks
\end{document}